\documentclass[preprint,12pt]{elsarticle}
\usepackage{lineno,hyperref}
\modulolinenumbers[5]

\journal{Computer Physics Communications}

\usepackage{lipsum}
\makeatletter
\def\ps@pprintTitle{%
 \let\@oddhead\@empty
 \let\@evenhead\@empty
 \def\@oddfoot{}%
 \let\@evenfoot\@oddfoot}
\makeatother

\usepackage[T1]{fontenc}
\usepackage[latin9]{inputenc}
\usepackage{url}
\usepackage{amsmath}
\usepackage{amssymb}
\usepackage{graphicx}
\usepackage{esint}
\usepackage{bbm}
\usepackage{tikz-cd}
\makeatletter

\newcommand{\GRdot}{.}

\usepackage{enumitem}		

\usepackage{color}
\usepackage{parskip}

\newcommand{\IdentOp}{\mathbbm{1}}

\global\long\def\ket#1{\mbox{\ensuremath{\left|#1\right\rangle }}}
\global\long\def\bra#1{\left\langle #1\right|}
\global\long\def\braket#1#2{\left\langle #1\vphantom{#2}\right|\left.#2\vphantom{#1}\right\rangle }
\global\long\def\ketbra#1#2{\left|#2\vphantom{#1}\right\rangle \left\langle #1\vphantom{#2}\right|}
\global\long\def\Braket#1#2#3{\left\langle #1\vphantom{\{#2#3\}}\right|#2\left|\vphantom{\{#1#2\}}#3\right\rangle }

\usepackage{setspace}

\makeatother

\begin{document}

\begin{frontmatter}

\title{Quantum Dynamics in Phase space using the Biorthogonal von Neumann bases: Algorithmic Considerations}

\author[xxx]{Shai Machnes\corref{cor1}}
\author[xxx]{Elie Ass{\'e}mat}
\author[xxx]{David Tannor}
\address[xxx]{Dept. of Chemical Physics, Weizmann Institute of Science, 76100 Rehovot, Israel}
\cortext[cor1]{Corresponding author, shai.machnes@gmail.com}

\begin{abstract}

The von Neumann lattice refers to a discrete basis of Gaussians located on a lattice in phase space.  It
provides an attractive approach for solving quantum mechanical problems, allowing the pruning of
tensor-product basis sets using phase space considerations. In a series of recent articles
Shimshovitz et al. [\emph{Phys. Rev. Lett.} \textbf{109} 7 (2012)],
Takemoto et al. [\emph{Journal of Chemical Physics} \textbf{137} 1 (2012)]
Machnes et al. [\emph{Journal of Chemical Physics}, accepted (2016)]), we have introduced two key new elements into the method: a formalism for converging the basis and for
efficient pruning by use of the biorthogonal basis.  In this paper we review the key
components of the theory and then present new, efficient and parallelizable iterative algorithms for
solving the time-independent and time-dependent Schrodinger equations. The algorithms dynamically
determine the active reduced basis iteratively without resorting to classical analogs. These
algorithmic developments, combined with the previous formal developments, allow quantum dynamics
to be performed directly and economically in phase space.  We provide two illustrative examples:
double-well tunneling and double ionization of helium.

\end{abstract}

\end{frontmatter}

\pagebreak
\addtocontents{toc}{\protect\setstretch{0.7}}
\tableofcontents{}

\section{Introduction}

The quantum description of multiparticle physical systems is typically characterized
by an extremely large Hilbert space, making any sort of wavefunction simulation extremely challenging \cite{KohnNobel}.
Examples range from electronic structure, ionization and high harmonic generation in multielectron systems to chemical reactions and photodissociation of polyatomic molecules. It is therefore crucial to find a compact representation of quantum states which admits explicit control of accuracy.

A promising direction is the class of methods in which states are
represented explicitly as phase space objects. Such methods
reduce resource usage to only those areas of phase space which are
actually occupied. Furthermore, such representations reveal
the physical nature of the state and its classical correspondence.
Despite some recent progress \cite{PollakWigner,RabitzEfficientWignerOpen}, the most commonly used phase space representation,
the Wigner representation, is not well suited for computational use.

To fill the need for a numerically accurate and efficient phase space
based approach, the \emph{periodic von Neumann} (PvN) representation
was developed \cite{PvB-1st,Asaf-PRL,Norio}. The method is based
on a lattice of phase space localized Gaussians as proposed by von
Neumann \cite{von-Neumann-grid-1} (see also Gabor \cite{von-Neumann-grid-3-Gabor}).
However, straightforward implementation of the von Neumann approach
is plagued with difficulties, due to the non-orthogonal nature of
the basis \cite{von-Neumann-grid-4-Davis-and-Heller}. The removal of these difficulties
relies on two key insights. First, that the poor convergence of the von Neumann lattice is a consequence
of it being a non-orthogonal basis.  By projecting the basis into the Hilbert space spanned by a discrete Fourier basis of the same size \cite{Fourier-grid-1,Fourier-grid-2,Fourier-grid-3},
we obtain exponential convergence for a wide range of functions.
Second, the recognition that although the von Neumann basis is phase space localized, the biorthogonal basis which we call BvN  (Biorthogonal von Neumann), is not.  Since the latter determines the coefficients, there is typically no sparsity at all in the representation. However, by exchanging the roles of the von Neumann basis and its biorthogonal basis, the representation typically becomes extremely sparse. This gives rise to the name of the method PvB (Periodic von Neumann with Biorthogonal exchange), and we will refer to the PvB representation.  With these two modifications the method
combines the best of both worlds: it is mathematically equivalent
to the Fourier grid, but has the flexibility that basis functions can be placed only in the areas of phase space where they are needed.

In \cite{The-Math-Paper} we developed the mathematical underpinnings of the method, in particular the projection onto subspaces spanned by non-orthogonal bases. In this manuscript we review the mathematical background and develop the algorithms necessary for an efficient implementation of the method.
These developments become crucial when applying the methodology to more challenging systems,
such as the helium atom in strong fields \cite{PvB-MCTDH-comparison-paper}.  The paper also contains additional formal developments, casting the PvB approach as a similarity transformation and drawing parallels with the Wigner and Husimi representations.

The paper is organized as follows. Section \ref{sec:Theory} introduces the formal theory. Section \ref{sec:Algorithms} deals with the algorithmic issues relating to the implementation of PvB. Some illustrative examples are given in Section \ref{sec:Examples}. Finally in Section \ref{sec:Conclusions-and-outlook} we discuss the outlook for the PvB method, and how some of its current limitations may be alleviated.

\section{\label{sec:Theory}Theory}

In this section we provide a brief summary of the main definitions, expressions and insights in the PvB method, in order to provide the context for the algorithms and implementation considerations discussed in Section. \ref{sec:Algorithms}.

We begin with a brief review of the PvB approach in Section \ref{sub:Review-of-PvB}.
Section \ref{sub:PvB-as-a-similarity-trans} recasts PvB as a similarity
transformation of the Fourier-grid projected von Neumann lattice. Section \ref{sub:The-reduced-basis} introduces the \emph{reduced
PvB representation}, allowing one to represent only those regions
of phase space actually occupied by the current state. In Section \ref{sub:The-TDSE-for-the-reduced_state} we derive the form of the Schr{\"o}dinger equations in the reduced representation.
These subjects are discussed in greater detail in \cite{The-Math-Paper}.

\subsection{\label{sub:Review-of-PvB}Review of PvB}

The PvB representation, developed in \cite{PvB-1st,Asaf-PRL,Norio},
is based on projecting the von Neumann lattice of phase space localized
Gaussians (Section \ref{sub:The-von-Neumann}) onto the Fourier grid
(Section \ref{sub:The-Fourier-Grid}). As the projected Gaussians
form a non-orthogonal basis, the representation of objects in this
Hilbert space requires the definition of two biorthogonal bases ---
one for the bras and the other for the kets (Section \ref{sub:Biorthogonal-bases}).
These elements are combined together in the PvB method (Section \ref{sub:PvN,PvB}).
Finally, the Schr{\"o}dinger equation is recast in the PvB representation (Section \ref{sub:Schroedinger-equation-in-PvB}).

\subsubsection{\label{sub:The-Fourier-Grid}The Fourier grid}

We briefly review the pseudospectral Fourier method \cite{FG-Kosloff,FG-2} (also known as the periodic sinc discrete variable representation (DVR \cite{FG-3})), which serves as the underlying Hilbert space on which we construct the PvB method. See also \cite{The-Math-Paper,David-Textbook}.

Functions with support on a finite segment $x\in\left[0,L\right]$, may be assumed, without loss of generality, to have cyclic boundary conditions. All such functions are spanned by
\begin{equation}
\varphi_{n}\left(x\right)=\frac{1}{\sqrt{L}}\exp\left(2\pi i\frac{x}{L}n\right)=\frac{1}{\sqrt{L}}\exp\left(ik_{n}x\right),\,\,\,\forall n\in\mathbb{Z},\label{eq:spectral-basis}
\end{equation}
with $k_{n}=\frac{2\pi}{L}n$. Limiting bandwidth to $K$ defines a rectangular area of phase space of area $2KL$,
which is spanned by the orthonormal \emph{spectral basis} of size $N=2n_{\textrm{max}}$,
$\left\{ \varphi_{n}\right\} _{n\in\left[-n_{\textrm{max}}+1,\ldots, n_{\textrm{max}}\right]}$, where
$n_{\textrm{max}}:=\left\lfloor \frac{KL}{2\pi}\right\rfloor.$ With an inner product defined as
$\left\langle f,g\right\rangle :=\int_{0}^{L}f^{*}\left(x\right)g\left(x\right)dx$,
this constitutes the \emph{Fourier grid (FG)} Hilbert space, $\mathcal{H}$.

Any $f\in\mathcal{H}$ may be expanded as
$f\left(x\right)=\sum_{n=1}^{N}\left\langle \varphi_{n},f\right\rangle \varphi_{n}\left(x\right)$. We may now represent $f$ by the
column vector of the expansion coefficients, $\vec{f}_{\varphi}$ $=$
$\left(\left\langle \varphi_{1},f\right\rangle ,\left\langle \varphi_{2},f\right\rangle ,\ldots\left\langle \varphi_{N},f\right\rangle \right)^{T}$. The norm
remains unchanged in the discrete representation, $\vec{f}_{\varphi}^{\dagger}\vec{g}_{\varphi}=\left\langle f,g\right\rangle$.

The FG Hilbert space may also be spanned by another basis, whose coefficients can be computed without integration.
Consider the set of \emph{Fourier grid points}, a set of $N$ equidistant sampling points,
$\left\{ x_{j}=x_{0}+L\frac{j}{N}\right\} _{j=0}^{N-1}$ for some $x_{0} \in \left[0,\frac{L}{N}\right]$.
Define the \emph{pseudo-spectral basis}, $\Theta=\left\{ \theta_{m}\left(x\right)\right\} _{m=1}^{N}$
as the set of orthogonal \emph{pseudo-spectral functions}, within
the FG Hilbert space, such that any function $f\left(x\right)\in\mathcal{H}$
can be expanded as
\begin{equation}
f\left(x\right)=\sum_{m=1}^{N}f\left(x_{m}\right)\theta_{m}\left(x\right).\label{eq:f_expanson_in_theta}
\end{equation}
where
\begin{equation}\label{eq:pseudo-spectral-ortho}
\theta_{m}\left(x\right)=\frac{1}{N}\sum_{n=-n_{\textrm{max}}+1}^{n_{\textrm{max}}}\exp\left(ik_{n}\left(x-x_{m}\right)\right)=
e^{i\pi\frac{x-x_{m}}{L}}\textrm{D}_\textrm{A}\left(N,2\pi\frac{x-x_{m}}{L}\right),
\end{equation}
with $\textrm{D}_\textrm{A}\left(N,\alpha\right):=\frac{\sin\left(N\frac{\alpha}{2}\right)}{N\sin\left(\frac{\alpha}{2}\right)}$
being the periodic sinc functions (also known as the Dirichlet functions), which are localized around $x_{m}$. $\alpha:=2\pi\frac{x-x_m}{L}$.
Note that $\left\langle \theta_{n},\theta_{m}\right\rangle =\frac{N}{L}\delta_{n,m}$. $f$ may now be represented by its \emph{sampling vector}, $\vec{f}=\vec{f}_{\theta}=\left(f\left(x_{1}\right),f\left(x_{2}\right)\ldots,f\left(x_{N}\right)\right)^{T}$, dropping the basis designation when implied by context.

Define the orthogonal projector into the FG Hilbert space by
\begin{equation}
\mathcal{P}:=\frac{L}{N}\sum_{m=1}^{N}\ketbra{\theta_{m}}{\theta_{m}}=\sum_{n=1}^{N}\ketbra{\varphi_{n}}{\varphi_{n}}.
\end{equation}
$\mathcal{P}$ minimizes the distance to the projected state \cite{Porat}. In contrast, for a function $f\left(x\right)\mathcal{\notin H}$, the projection $\mathcal{P}_\mathcal{S}f\left(x\right):=\sum_{j}f\left(x_{j}\right)\theta_{j}\left(x\right)$
is not an orthogonal projection, but rather
the \emph{sampling pseudo-projection}, or \emph{collocation projection}.

The FG and its associated spectral and pseudo-spectral bases are just one possible choice for the phase space underlying PvB.
See  \cite{Asaf-Review,Asaf-non-FG,TuckerCarrington} for alternative possibilities.

\subsubsection{\label{sub:The-von-Neumann}The von Neumann lattice and its projection onto the Fourier grid (PvN)}

Consider the infinite \emph{von Neumann lattice} \cite{von-Neumann-grid-1,von-Neumann-grid-2,von-Neumann-grid-3-Gabor} of Gaussians in the $\left(x,p\right)$ plane. Let $\left(\bar{x}_{i},\bar{p}_{i}\right)$ indicate the center of Gaussian $i$, and let $\left(\Delta x,\Delta p\right)$ be the spacing between the lattice sites. We define
\begin{equation}
g_{\bar{x}_i,\bar{p}_i}(x):=\left(\frac{1}{2\pi\sigma_{x}^{2}}\right)^{1/4}\exp\left(-\left(\frac{x-\bar{x}_i}{2\sigma_{x}}\right)^{2}+\frac{i}{\hbar}\bar{p}_i(x-\bar{x}_i)\right).
\end{equation}

 The representation of a state as a linear combination of the above Gaussians converges poorly \cite{von-Neumann-grid-4-Davis-and-Heller,Daubechies}. To overcome this problem, we truncate the von Neumann lattice to a finite domain of size $N_{x}\times N_{p}=N$, project the Gaussians onto the FG, and reformulate them in a cyclic fashion,
\begin{equation}\label{eq:PvN-basis}
g_{\bar{x}_i,\bar{p}_i}^\textrm{mod}(x):= \left(\mathcal{Q}g_{0,0}(\textrm{mod}_{L}\left(x-\bar{x}_i\right)\right)e^{\frac{i}{\hbar}\bar{p}_i\,\textrm{mod}_{L}\left(x-\bar{x}_i\right)}.
\end{equation}
with $\textrm{mod}_{L}x:=x-L\left\lfloor \frac{x}{L}\right\rfloor$. Thus, we achieve convergence properties equivalent to the FG \cite{PvB-1st}. The cyclic projected Gaussians, known as the \emph{Periodic von Neumann} or \emph{PvN basis}, are \emph{periodic Gabor functions}, admit a generalized version of the Fast Fourier Transform  \cite{Fast-Gabor-transform}. For the conditioning of the overlap matrices (eq. \ref{eq:Overlap-matrices}), it is beneficial to choose the von Neumann lattice points such that $\bar{x}_i$ are a subset of the FG sample points and  $\bar{p}_i$  are a subset of the spectral basis frequencies.

The PvN basis in the $\Theta$ representation is
\begin{equation}
G_{jk}:=g_{\bar{x}_{k},\bar{p}_{k}}^{\textrm{mod}}(x_{j}).\label{eq:G_jk},
\end{equation}
or in Dirac notation:
\begin{equation}
\ket{g_{\bar{x}_{k},\bar{p}_{k}}^{\textrm{mod}}}=
\sum_{m=1}^{N}\ket{x} \braket{\theta_{m}}{g_{\bar{x}_{k},\bar{p}_{k}}} \approx
\sum_{m=1}^{N}\ket{\theta_{m}} \braket{\theta_{m}}{g_{\bar{x}_{k},\bar{p}_{k}}}.
\end{equation}

The FG defines an area of $\left(2K\right)L=2\pi N$ in phase space. Therefore one may informally assign a phase space area of $2\pi$ to each of the $N$ Gaussians of the PvN basis, and consider them \textquotedblleft phase space pixels\textquotedblright{}. Defining $P:=\hbar K$, each such pixel covers a phase space area of $2 \pi \hbar =h$.

\subsubsection{\label{sub:Biorthogonal-bases}Biorthogonal bases}

In this section we consider the general theory of non-orthogonal bases and their biorthogonal bases. In the next section we will specialize to the periodic von Neumann basis and its biorthogonal basis. In anticipation of that section, we use the notation $G$ and $B$ here.

Any set of $N$ linearly independent vectors $\mathcal{G}=\left\{ \ket{g_{k}}\right\} _{k=1}^{N}$ in $\mathcal{H}$ may serve as a non orthogonal basis of $\mathcal{H}$. Let $\mathcal{G}$ be represented in the $\Theta$ orthogonal basis of $\mathcal{H}$ by the invertible matrix $G$. Let $\mathcal{B}$ be a similarly defined non-orthogonal basis of $\mathcal{H}$, represented in $\Theta$ by $B$. The bases $\mathcal{G}$ and $\mathcal{B}$ are considered \emph{biorthogonal bases} (a reciprocal relationship) if
\begin{equation}
G^{\dagger}B={\IdentOp}_{N}\,\,\,\Longleftrightarrow\,\,\,\braket{g_{k}}{b_{j}}=\delta_{kj}.\label{eq:bi-otrho prelim}
\end{equation}
$G$ is the left pseudo-inverse of $B$ (and vice versa)
\begin{equation}\label{eq:B_of_G_of_B}
G^{\dagger}=\left(B^\dagger B\right)^{-1}B^\dagger\,\,\,\Longleftrightarrow\,\,\,B=G\left(G^\dagger G\right)^{-1}.
\end{equation}
One may show that $G$ and $B$ also satisfy the completeness relation
\begin{equation}
\begin{array}{ccc}
BG^{\dagger}={\IdentOp}_{N} & \quad\textrm{\ensuremath{\Longleftrightarrow}\quad} & \sum_{j}\ket{b_{j}}\bra{g_{j}}={\IdentOp}.\end{array}\label{eq:GB completeness}
\end{equation}
Define $S$ and $S^{-1}$ as the \emph{overlap matrices} of the $\mathcal{G}$ and $\mathcal{B}$ bases, respectively,
\begin{equation}
\begin{array}{rcccrclcrcl}
S & := & G^{\dagger}G, &  & G & = & BS, &  & \vec{\psi}_{B} & = & S\vec{\psi}_{G},\\
S^{-1} & = & B^{\dagger}B, &  & B & = & GS^{-1}, &  & \vec{\psi}_{G} & := & S^{-1}\vec{\psi}_{B}.
\end{array}\label{eq:Overlap-matrices}
\end{equation}

Any $\ket{\psi}\in\mathcal{H}$, with $\vec{\psi}$ being its pseudo-spectral representation, may be expressed in $\mathcal{B}$ as
\begin{equation}
\vec{\psi}_{B}:=G^{\dagger}\vec{\psi},\label{eq:psi_B}
\end{equation}
with $\psi=B\vec{\psi_B}$. \emph{One may therefore consider $G^{\dagger}$ to represent the bras $\left\{ \bra{g_{k}}\right\} $, while $B$ represents the kets, $\left\{ \ket{b_{j}}\right\} $}. Norms of vectors in $B$ and $G$ are computed via $\left\Vert \psi_{G}\right\Vert ^{2}=\vec{\psi^{\dagger}}_{G}S\vec{\psi}_{G}$ and $\left\Vert \psi_{B}\right\Vert ^{2}=\vec{\psi^{\dagger}}_{B}S^{-1}\vec{\psi}_{B}$. Note that the $G$ and $B$ bases cannot be independently normalized.

\subsubsection{\label{sub:PvN,PvB}BvN - The basis biorthogonal to PvN}

Many wavefunctions of interest, whether bound states or traveling
wavepackets, are fairly well localized in phase space. Therefore
a representation whose coefficients are $\beta_{k}=\braket{g_{k}}{\psi}$,
is expected to be sparse, i.e. to have many near-zero elements.

Any state $\ket{\psi}\in\mathcal{H}$ can be represented in the \emph{Periodic von Neumann} (PvN) basis as
\begin{equation}\label{eq:psi_expanded_in_G}
\ket{\psi}=\left(\sum_{j=1}^{N}\ket{g_{j}}\bra{b_{j}}\right)\ket{\psi}=\sum_{j=1}^{N}\braket{b_{j}}{\psi}\ket{g_{j}}.
\end{equation}
Alternatively, it may be represented in the BvN basis as
\begin{equation}\label{eq:psi_expanded_in_B}
\ket{\psi}=\left(\sum_{j=1}^{N}\ket{b_{j}}\bra{g_{j}}\right)\ket{\psi}=\sum_{j=1}^{N}\braket{g_{j}}{\psi}\ket{b_{j}}.
\end{equation}
The PvN and BvN bases span exactly the same Hilbert space as the spectral and pseudospectral bases and contain the identical number of functions. Therefore there is no ambiguity or redundancy, and no loss of information.

\begin{figure}
\noindent \begin{centering}
\includegraphics[scale=0.6]{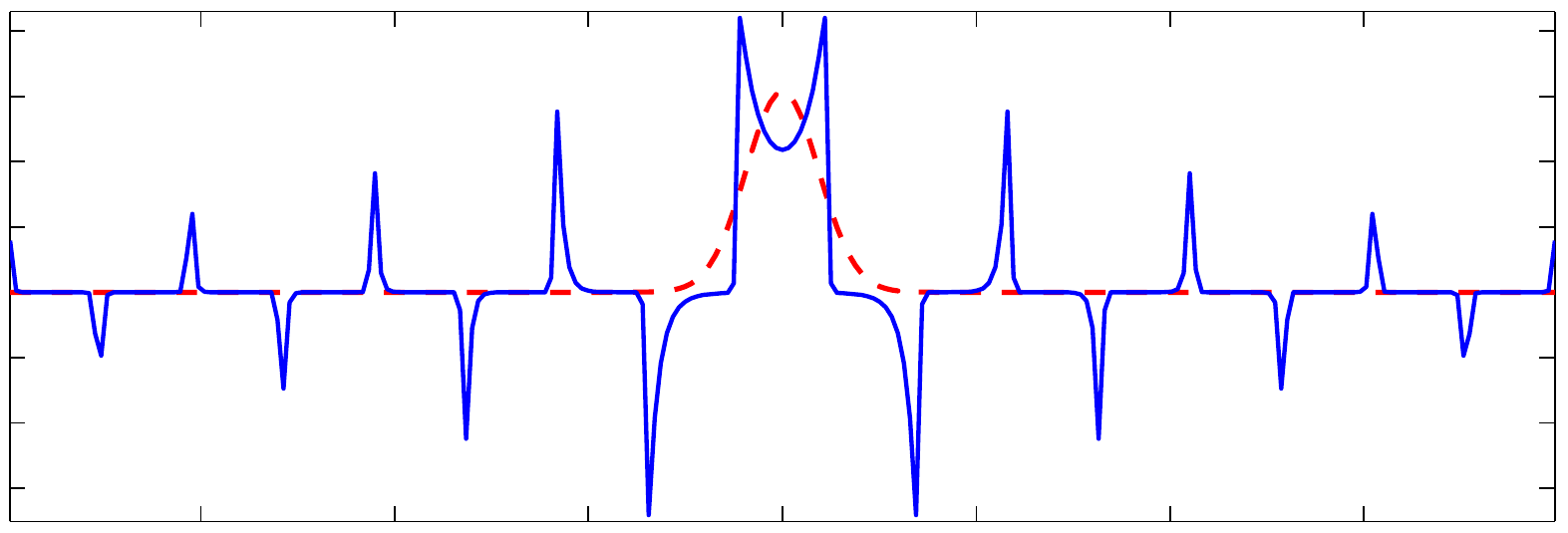}
\par\end{centering}

\noindent \begin{centering}
\protect\caption{\label{fig:b-function}Gaussian function (red dashed) and its biorthogonal
$b$ function (blue). Note that $b$ is highly non-local, non-smooth and non-monotonic.}

\par\end{centering}

\end{figure}

\begin{figure}
\noindent \begin{centering}
\includegraphics[scale=0.3888]{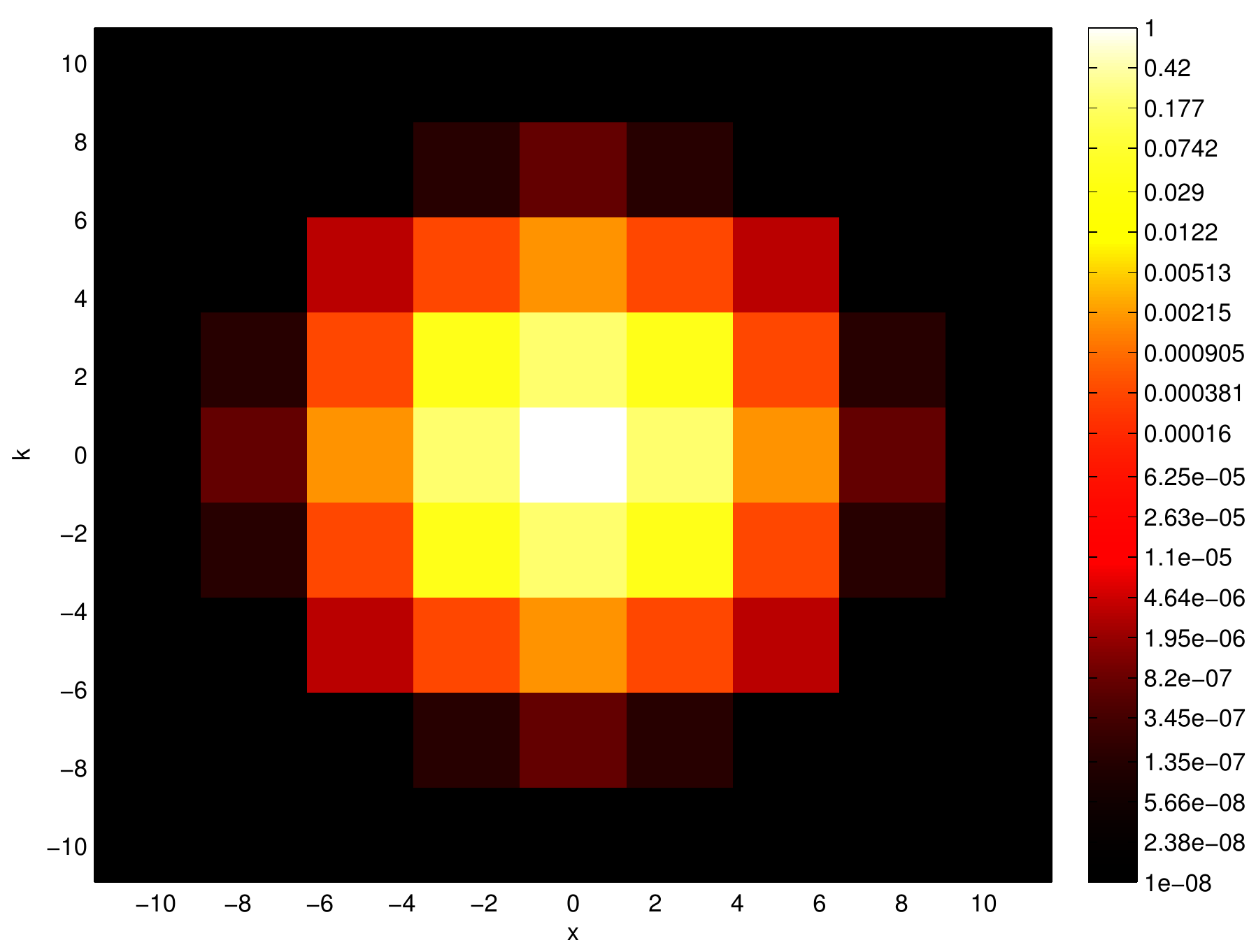}\includegraphics[scale=0.4188]{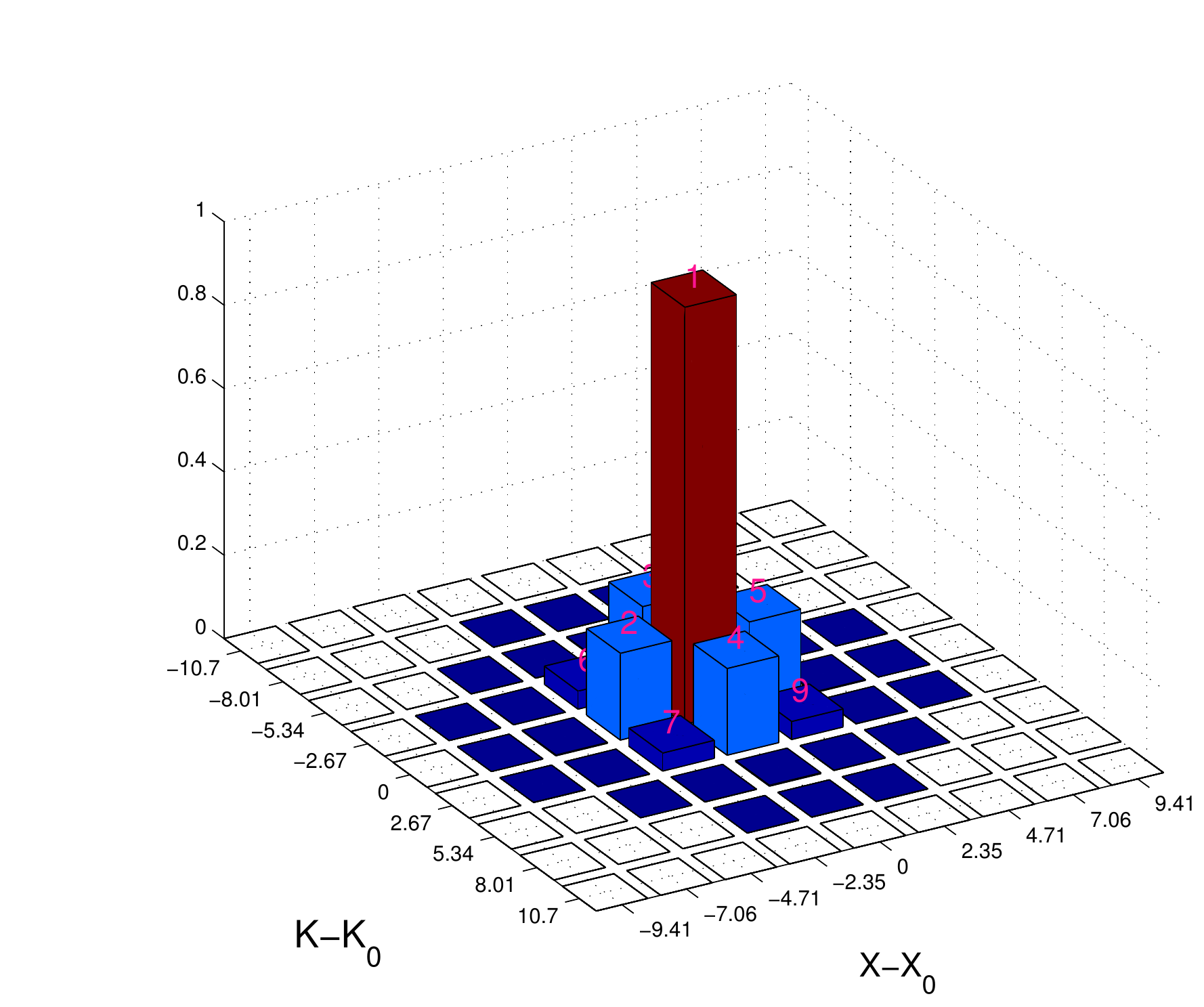}\\
\includegraphics[scale=0.4551]{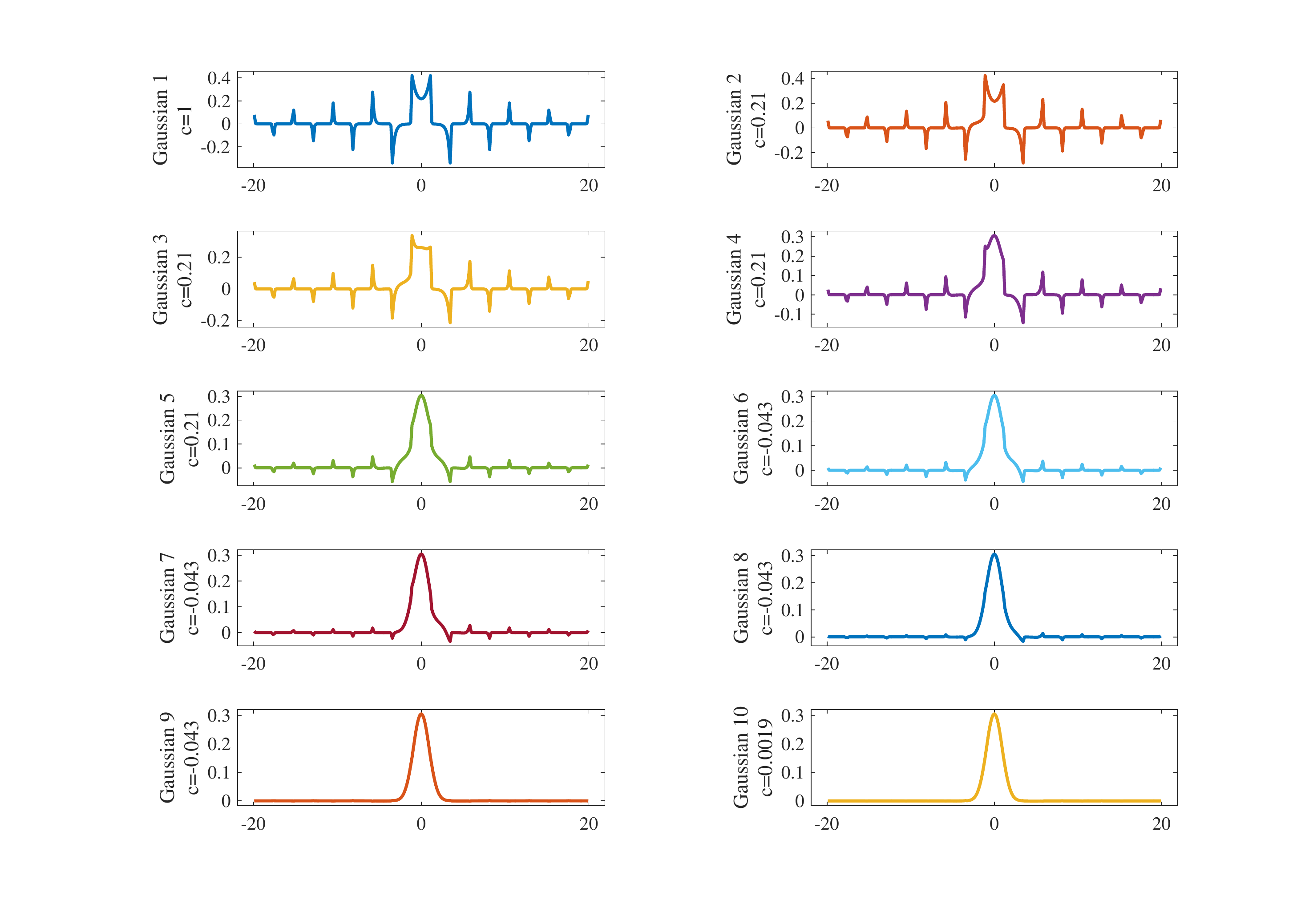}
\par\end{centering}

\noindent \centering{}\protect\caption{\label{fig:State-buildup-by-Bs}Top left: a Gaussian in the PvB representation shown as a heat map (logarithmic color scale). Top right: the same as left but shown as a bar plot. The value plotted in each cell of the von Neumann lattice is the absolute value of the overlap of the state with the Gaussian centered at that cell of the lattice,
$\left|{\left\langle{g_{\bar{x}_{j},\bar{k}_{j}}}\right|\left.\psi\right\rangle}\right|$.
The overlap of neighboring Gaussians is non-vanishing, as is apparent from the amplitude of the bars $2\ldots5$
(purple in the top right plot). The bottom plots detail how the Gaussian wavefunction is accumulated from weighted $\ket{b}$ functions. Note the counterintuitive result that a smooth Gaussian is expressed by a small sum of non-smooth non-local functions. }
\end{figure}

Note that although the $\ket{g}$ states are localized in phase space, the $\ket{b}$ states are highly delocalized, as can be seen in fig. \ref{fig:b-function}. As a result, the representation in the BvN basis, eq. \ref{eq:psi_expanded_in_B}, is sparse since the localized $\bra{g}$ bras determine the coefficients, while the representation in the PvN representation, eq. \ref{eq:psi_expanded_in_G}, is not sparse since the delocalized $\bra{b}$ bras determine the coefficients. See fig. \ref{fig:State-buildup-by-Bs} for an illustrative example.

From this point onward, we shall assume all states are in $\mathcal{H}$ and use the discrete $\Theta$ representation.
We will therefore drop the explicit vector notation, e.g. in the states $\vec{\psi}$ and $\vec{\psi}_{B}$.

\subsubsection{\label{sub:Schroedinger-equation-in-PvB}The Schr{\"o}dinger equation in PvB}

In the PvB representation, the time independent and time dependent Schr{\"o}dinger equations
(TISE and TDSE, respectively) take the forms,
\begin{equation}
\left(G^{\dagger}HB\right)\psi_{B}=\lambda\psi_{B}. \label{eq:TISE_in_B}
\end{equation}
and
\begin{equation}
\partial_{t}\psi_{B}=-\frac{i}{\hbar}\left(G^{\dagger}HB\right)\psi_{B}.\label{eq:TDSE_in_B}
\end{equation}
The term $G^{\dagger}HB=B^{-1}HB$ is a similarity transformation of $H$ and therefore all eigenvalues are real and all evolutions are unitary. The TISE, eq. \ref{eq:TISE_in_B}, may be rephrased as a generalized eigenvalue problem,
\begin{equation}
\left(B^{\dagger}HB\right)\psi_{B}=\left(B^{\dagger}B\right)\lambda\psi_{B}.
\end{equation}

One may rewrite the TDSE in four distinct ways, all strictly equivalent. We denote a Hamiltonian taking a state in basis $X$ and returning a state in basis $Y$ by $H_{YX}$,
\begin{equation}
\begin{array}{ccccccccccc}
G^{\dagger}HB & = & B^{-1}HB & =: & H_{BB} &  & \longrightarrow &  & \partial_{t}\psi_{B} & = & -\frac{i}{\hbar}\,H_{BB}\,\psi_{B\,},\\
B^{\dagger}HB & = & G^{-1}HB & =: & H_{GB} &  & \longrightarrow &  & \partial_{t}\psi_{B} & = & -\frac{i}{\hbar}\,S\,H_{GB}\,\psi_{B}\,,\\
G^{\dagger}HG & = & B^{-1}HG & =: & H_{BG} &  & \longrightarrow &  & \partial_{t}\psi_{B} & = & -\frac{i}{\hbar}\,H_{BG}\,S^{-1}\,\psi_{B}\,,\\
B^{\dagger}HG & = & G^{-1}HG & =: & H_{GG} &  & \longrightarrow &  & \partial_{t}\psi_{B} & = & -\frac{i}{\hbar}\,S\,H_{GG}\,S^{-1}\,\psi_{B}\,.
\end{array}\label{eq:H_GB_BB_BG_GG}
\end{equation}
Although the forms are mathematically equivalent, they require different computational efforts. For example, $G^{\dagger}HG$ is quick to compute as the integrations on both sides are with the highly-local Gaussians. This is counter-balanced by the need to compute $S^{-1}$.

\subsection{\label{sub:PvB-as-a-similarity-trans}PvB as a
similarity transformation }

PvB may be reformulated as a similarity transformation from the pseudospectral basis to the non-orthogonal $B$ basis, without
explicitly referring to the $G$ basis. This allows a natural extension of PvB to density matrices and operators, which in turn clarifies
the connection between PvB and the discrete versions of the Husimi and Wigner representations.

\subsubsection{PvB expressed entirely in terms of $B$}

The basis $\{\ket{b_j}\}_{j=1}^{N}$ is linearly independent and hence the $B$ matrix is invertible (see discussion in Section \ref{sub:Biorthogonal-bases}). Therefore, any state in the FG may be converted to and from the $B$ representation by

\begin{equation}
\ket{\psi_{B}}=B^{-1}\ket{\psi}\,\,\,\Longleftrightarrow\,\,\,\ket{\psi}=B\ket{\psi_{B}}.\label{eq:psi_B_ket_similarity}
\end{equation}

Requiring $\braket{\psi_{B}}{\psi_{B}}=1$, and considering eq. \ref{eq:psi_B_ket_similarity}
suggests that the bra transforms as

\begin{equation}
\bra{\psi_{B}}=\bra{\psi}B.\label{eq:psi_B_bra_similarity}
\end{equation}

Substituting eq. \ref{eq:psi_B_ket_similarity} into the TDSE and multiplying on the left by $B^{-1}$, we arrive at
$i\hbar\partial_{t}\ket{\psi_{B}}=B^{-1}HB\ket{\psi_{B}}$. This indicates that the Hamiltonian is converted to the $B$ basis via a similarity transform

\begin{equation}
H_{BB}=B^{-1}HB,\,\,\,\,\,\,\,i\hbar\partial_{t}\ket{\psi_{B}}=H_{BB}\ket{\psi_{B}}.
\end{equation}

Any density matrix $\rho$ may be expanded as $\rho=\sum_{k}{p_{k}}\ket{\psi_{k}}\bra{\psi_{k}}$. Utilizing eq. \ref{eq:psi_B_ket_similarity} and \ref{eq:psi_B_bra_similarity}, $\rho$ transforms via $\rho=\sum_{k}\left(B\ket{\psi_{B,k}}\right)\left(\bra{\psi_{B,k}}B^{-1}\right)=B\rho_{BB}B^{-1}$, giving

\begin{equation}
\rho_{BB}=B^{-1}\rho B.\label{eq:dm-in-pvb}
\end{equation}

The equation of motion remains unchanged, i.e. $i\hbar\partial_{t}\rho_{BB}=\left[H_{B},\rho_{BB}\right]$.
Other operators transform similarly, $\mathcal{O}_{B}=B^{-1}\mathcal{O}B$, as do their respective equations of motion.

\subsubsection{\label{sub:Connection-to-Husimi}Connection to the Husimi and Wigner
representations}

The Husimi phase space representation of the density matrix at location
$\left(x_{0},p_{0}\right)$ is defined as \cite{Husimi}.

\begin{equation}
\mathcal{Q}\left(x_{0},p_{0}\right)=\Braket{g_{x_{0},p_{0}}}{\rho}{g_{x_{0},p_{0}}}.\label{eq:Husimi_Q}
\end{equation}

The Husimi representation may alternatively be defined as a Gaussian
filter applied to the Wigner representation \cite{GaussianWignerHusimi}:
\begin{equation}
\mathcal{Q}\left(x_{0},p_{0}\right)=\int\int\mathcal{W}\left(y_{0},q_{0}\right)e^{-2\left(\left(x_{0}-y_{0}\right)^{2}+\left(p_{0}-q_{0}\right)^{2}\right)}dy_{0}dq_{0}.
\end{equation}
Somewhat counterintuitively, despite this Gaussian filter it can be shown to be informationally equivalent to the Wigner representation  \cite{Wigner-1,Wigner-2}.

Both the Husimi and Wigner representations are complete and consistent representations
of quantum mechanics  \cite{Wigner-Husimi-equiv-to-standard-1,Wigner-Husimi-equiv-to-standard-2}.
Expanding on ideas in \cite{David-on-Husimi} we observe the following: from eqs. \ref{eq:G_jk} and \ref{eq:psi_B}, we know that an element of the vector $\psi_B$ associated with the Gaussian centered
at $\left(\bar{x}_k,\bar{p}_k\right)$ is $\braket{g_{\bar{x}_k,\bar{p}_k}}{\psi}$. For a
pure state density matrix, using eq. \ref{eq:dm-in-pvb} $\rho_{BB}=\ket{\psi_B}\bra{\psi_B}$, and therefore the $\left(k,k\right)$ element
of the density matrix will be $\left|\braket{g_{\bar{x}_k,\bar{p}_k}}{\psi}\right|^2$.
This is identical to the value of the Husimi representation of $\psi$ at this point, eq. \ref{eq:Husimi_Q}.
Thus, the Husimi representation can be viewed as the intensity associated with $\ket{\psi_{B}}$, defined not only at the von Neumann lattice points, but for continuous values of $x$ and $p$.

For arbitrary density matrices, PvB may be rewritten as the discrete
version of eq. \ref{eq:dm-in-pvb}
\begin{equation}
\rho_{BB}=\Braket{g_{x_{i},p_{j}}}{\rho}{b_{x_{k},p_{l}}}.\label{eq:rho_B}
\end{equation}

Note the difference from eq. \ref{eq:Husimi_Q}. In light of the discussion
in Section \ref{sub:Schroedinger-equation-in-PvB},
we may view $\rho_{BB}$ (eq. \ref{eq:rho_B}) as similar in structure
to $G^{\dagger}HB$ (eq. \ref{eq:H_GB_BB_BG_GG}), whereas
Husimi's $Q$ (eq. \ref{eq:Husimi_Q}) is similar to $G^{\dagger}HG$
(again, eq. \ref{eq:H_GB_BB_BG_GG}).

\subsection{\label{sub:The-reduced-basis}The reduced Hilbert space}

\subsubsection{\label{sub:Definition_of_reduced}Defining the reduced Hilbert space, $\widetilde{\mathcal{H}}$}

As discussed above, if the state $\ket{\psi}_B$  that we are trying to represent is localized in phase space, it will have significant overlap with only a small number of the localized Gaussians. As a result, many of the coefficients of $\ket{\psi}_B$ will be negligible. We therefore define a reduced Hilbert subspace, $\widetilde{B}$, spanned by a subset of the $\ket{b}$ vectors, whose coefficients are all above some predefined threshold. In Section \ref{sub:Biorthogonal-bases-for} we define the biorthogonal basis for this reduced space, $\widetilde{G}$,  and in Section \ref{sub:The-Projector} the projector into this space.

Consider a Hilbert space of dimension $N$ spanned by the  orthogonal pseudospectral basis $\Theta$ (eq. \ref{eq:f_expanson_in_theta}) and a set of biorthogonal bases $\mathcal{B}$, $\mathcal{G}$, represented in $\Theta$ by the matrices, $B$ and $G$ (eq. \ref{eq:G_jk},\ref{eq:B_of_G_of_B}). For the sake of notational convenience, let us assume the first $\widetilde{N}$ coefficients in $\psi_{B}$ are significant,
while the remaining $N-\widetilde{N}$ are negligible. In such a case,
we can save computational resources by reducing the vector $\psi_{B}$
of length $N$ to a vector $\psi_{\widetilde{B}}$ of length $\widetilde{N}\ll N$.

Define the \emph{reduced subspace}, $\widetilde{\mathcal{H}}\subseteq\mathcal{H}$
as the Hilbert space spanned by the first $\widetilde{N}$ vectors of $B$, $\widetilde{B}$,
\begin{equation}
\mathcal{\widetilde{H}}=\textrm{span}\left(\left\{ \ket{b_{j}}\right\} _{j=1}^{\widetilde{N}}\right).\label{eq:red_subspace_def}
\end{equation}
Define \emph{the complementary subspace},$\bar{\mathcal{H}}$, by
\begin{equation}\label{eq:H-straight-sum}
\mathcal{H}=\widetilde{\mathcal{H}}\oplus\bar{\mathcal{H}}\,\,,
\end{equation}
requiring that \emph{every vector in the complementary subspace $\mathcal{\bar{H}}$
be orthogonal to every vector in the reduced subspace, $\widetilde{\mathcal{H}}$}.
As the $B$ basis is non-orthogonal, there is no partitioning of the $B$ functions that spans both subspaces. However, the complementary space may be spanned by $\bar{\mathcal{H}}=\textrm{span}\left(\left\{ \ket{g_{k}}\right\} _{k=\widetilde{N}+1}^{N}\right)$. Given any state $\ket{\psi}\in\mathcal{H}$, we may therefore decompose $\ket{\psi}=\ket{\widetilde{\psi}}+\ket{\bar{\psi}}$ s.t. $\braket{\widetilde{\psi}}{\bar{\psi}}=0$, and $\ket{\widetilde{\psi}}\in\widetilde{\mathcal{H}}$, $\ket{\bar{\psi}}\in\mathcal{\bar{H}},$
where $\ket{\widetilde{\psi}}$ is the \emph{reduced state} and $\ket{\bar{\psi}}$ \emph{the complementary state}.

\subsubsection{\label{sub:Biorthogonal-bases-for}Biorthogonal bases for $\widetilde{\mathcal{H}}$}

We now turn to the biorthogonal bases for the reduced subspace. Let us denote the first $\widetilde{N}$ columns of the $B$ matrix with the matrix $\widetilde{B}_{N\times\widetilde{N}}$, i.e. $\widetilde{B}$ is the $\Theta$ representation of the basis defining $\widetilde{H}$. $\widetilde{G}$ will be defined as a basis that is biorthogonal to
$\widetilde{B}$, and spans the same Hilbert space as $\widetilde{B}$. In \cite{The-Math-Paper} we show that both requirements are satisfied by the right pseudo-inverse
of $\widetilde{B}^\dagger$,
\begin{equation}
\widetilde{G}:=\widetilde{B}\left(\widetilde{B}^{\dagger}\widetilde{B}\right)^{-1},\label{eq:G_tilde_def}
\end{equation}
We stress that $\widetilde{B}$ is the natural basis of  $\widetilde{\mathcal{H}}$,
while $\widetilde{G}$ is defined in terms of $\widetilde{B}$. As will be seen in the following section, the columns of $\widetilde{G}$ are no longer exactly the projected Gaussians, but have been deformed as the result of the basis reduction. The overlap matrices in the reduced subspace are
\begin{equation}\label{eq:tilde_S_and_invS}
\widetilde{S}:=\left(\widetilde{B}^{\dagger}\widetilde{B}\right)^{-1}=\widetilde{G}^{\dagger}\widetilde{G},
~~~~\widetilde{S}^{-1}=\widetilde{B}^{\dagger}\widetilde{B}=\left(\widetilde{G}^{\dagger}\widetilde{G}\right)^{-1},
\end{equation}
from which it follows that $\widetilde{G}=\widetilde{B}\widetilde{S}$ and $\widetilde{B}=\widetilde{G}{{\widetilde{S}}^{-1}}$.

 We may also define the basis for the complementary Hilbert space $\bar{\mathcal{H}}$ with $\bar{G}_{N\times\left(N-\widetilde{N}\right)}$ as the last $N-\widetilde{N}$
columns of $G$, and $\bar{B}:=\bar{G}\left(\bar{G}^{\dagger}\bar{G}\right)^{-1}$. Note that here $\bar{G}$, not $\bar{B}$, is the natural basis. This is because the functions in $\bar{G}$, not $\bar{B}$, are orthogonal to the functions in $\widetilde{B}$.

\subsubsection{\label{sub:The-Projector}Projecting into $\widetilde{\mathcal{H}}$}

Consider the projector from the unreduced $\mathcal{H}$ to the reduced $\mathcal{\widetilde{H}}$.  This may be written in two alternate forms:
\begin{equation}\label{eq:P_Dirac}
\mathcal{\widetilde{P}}:=\sum_{j=1}^{\widetilde{N}}\ketbra{\widetilde{g}_{j}}{\widetilde{b}_{j}}=\sum_{j=1}^{\widetilde{N}}\ketbra{\widetilde{b}_{j}}{\widetilde{g}_{j}}\,\,.
\end{equation}
In the $\Theta$ representation (i.e. both input and output are represented on the Fourier grid, $\mathcal{\widetilde{P}}$ is given by the idempotent matrix
\begin{equation}
\widetilde{P}:=\widetilde{B}\widetilde{G}^{\dagger}.\label{eq:P_in_FG}
\end{equation}

\begin{figure}
\noindent \begin{centering}
\includegraphics[scale=0.6]{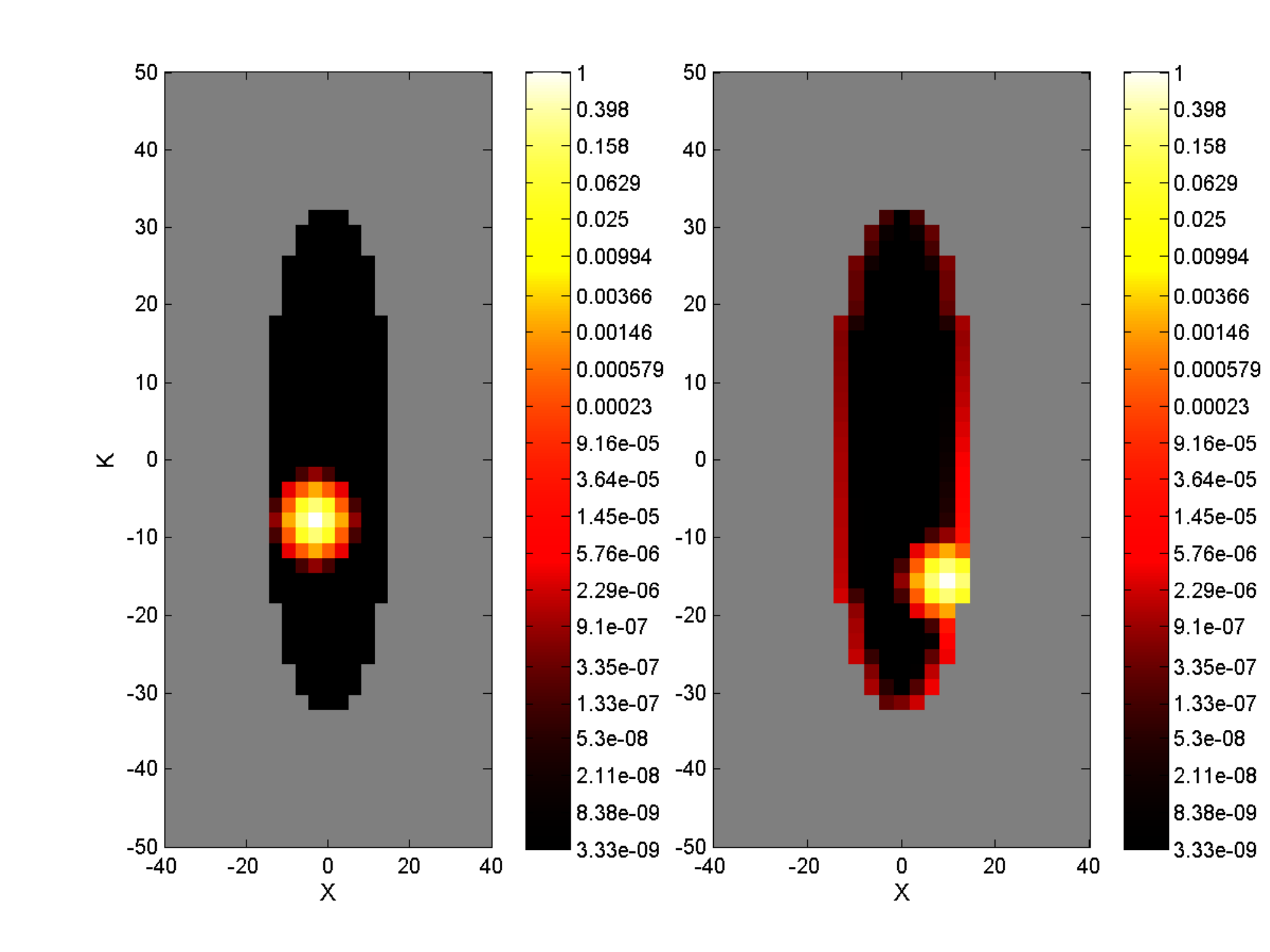}
\par\end{centering}
\noindent \centering{}\protect\caption{Depiction of modified Gaussians associated with the reduced basis. The reduced basis is the non-gray area in both plots. On the left, the modified Gaussian $\widetilde{g}$, which is centered around a non-edge location in the phase space, is almost unchanged. On the right, we see a heavily deformed Gaussian, whose center is close to the reduced subspace boundary. The $\tilde{G}$ basis is modified by a projection of the $G$ functions into the subspace spanned by $\tilde{B}$. The states are plotted as heat maps, where the value of each cell of the von Neumann lattice is the absolute value of the overlap of the state plotted (here the modified Gaussians), with the Gaussian centered at that cell of the lattice, $\left|{\left\langle{g_{\bar{x}_{j},\bar{k}_{j}}}\right|\left.\widetilde{g}\right\rangle}\right|$.\label{fig:deformed_Gaussians}}
\end{figure}

In \cite{The-Math-Paper} we showed that the $\widetilde{G}$ vectors spanning the reduced basis are the projection of the $G$ vectors into the reduced subspace, and
\begin{equation}\label{eq:tildeP_subtractive_on_gk}
\ket{\widetilde{g}_k}=\ket{{g}_k} - \sum_{j=\widetilde{N}+1}^{N} \braket{g_j}{g_k} \ket{\bar{b}_j}\,\,\,\forall k\le\widetilde{N}.
\end{equation}
This provides insight into the deformation of the Gaussians due to the basis reduction, as depicted in fig. \ref{fig:deformed_Gaussians}. The deformation is the result of the subtraction of $\ket{\bar{b}}$ vectors, weighted by the overlap of the Gaussians inside the reduced subspace with the Gaussians spanning the complementary space. This clarifies why Gaussians at the boundary of the reduced subspace are modified more --- they have a larger overlap with Gaussians outside the reduced subspace.

When projecting from the full to the reduced Hilbert space, the projector takes different forms depending on the source and target representations.  Denote by $\widetilde{P}_{YX}$ the projection operator from the full Hilbert space $\mathcal{H}$ into the reduced subspace $\widetilde{\mathcal{H}}$, where the Fourier grid representation of the input basis for $\mathcal{H}$ is the matrix $X$, and the Fourier grid representation for $\widetilde{\mathcal{H}}$ is the matrix $Y$. The projection operator transforms as do all operators, using $P_{YX}=Y^{-1}PX$  (when $Y$ is not square, $Y^{-1}$ should be taken to be the left pseudo-inverse of $Y$). For example, using $\widetilde{B}=BR$, the projector $\widetilde{P}_{\widetilde{G}G}$ takes the form $R^{\dagger}$, with
\begin{equation}
R_{N\times\widetilde{N}}:=\left(\begin{array}{cccc}
1 & 0 & 0 & \ldots\\
0 & 1 & \ddots & \ddots\\
\vdots & \ddots & \ddots & 0\\
0 & \ldots & 0 & 1\\
0 & 0 & \ldots & 0\\
0 & \ddots &  & 0\\
\vdots &  & \ddots & \vdots\\
0 & 0 & \ldots & 0
\end{array}\right)\label{eq:R_def}
\end{equation}

\begin{figure}
\vspace*{-1in}
\centerline{\begin{centering}\includegraphics[scale=0.6173]{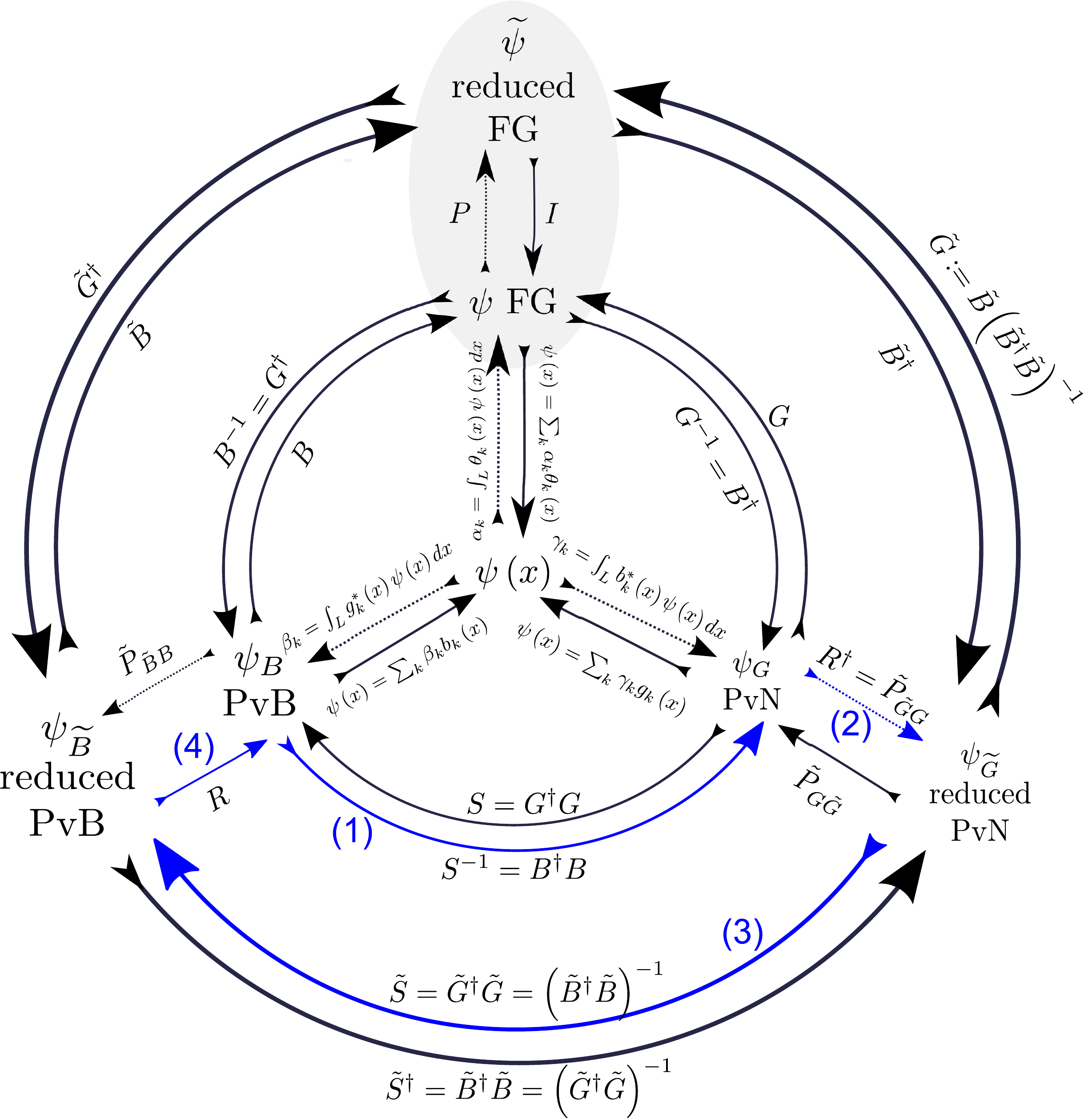}\end{centering}}
\noindent \centering{}\protect\caption{\label{fig:PEACE_sign}
Transformations and projections from continuous wavefunctions, $\psi(x)$ (center), to the Hilbert space spanned by the Fourier grid (inner circle) and the reduced subspace (outer circle). Dashed arrows indicate a projection (surjective; with information loss), continuous arrows
are injective (one-to-one but not necessarily on-to) mappings.
In the PvN representation, states are described as a sum of $g$'s, which serve as the kets, while $b$'s serve as the bras. In the PvB representation the roles are reversed.
The nomenclature $\widetilde{P}_{YX}$ is used to indicate the projection
from representation $X$ of $\mathcal{H}$ into representation
$Y$ of $\widetilde{\mathcal{H}}$. This diagram may be used, to construct $\widetilde{P}_{\widetilde{B}B}$ for example, by
going through the following series of transformations: $\left(1\right)$, from the $B$ basis
to $G$ using the transformation matrix $S^{-1}$, $\left(2\right)$ projecting
with $R^{\dagger}$, $\left(3\right)$ transforming back to $\widetilde{B}$ using $\widetilde{S}$,
$\left(4\right)$ transforming to base $B$ using $R$. Multiplying the matrices appearing from right to left gives
$\widetilde{P}_{BB}=R\widetilde{S}R^{\dagger}S^{-1}$.
Note that although we return to the same point from which started, the resulting operation is not equal to $1$ because of the loss of information in the projection in step $\left(2\right)$.}
\end{figure}

\cite{The-Math-Paper}. Figure \ref{fig:PEACE_sign} shows graphically how the transformation between these different representations can be combined in a wide variety of different ways.

\subsection{\label{sub:The-TDSE-for-the-reduced_state}The Hamiltonian and the Schr{\"o}dinger equations for the reduced state}

The Schr{\"o}dinger equation for the reduced state requires projecting the Hamiltonian into the reduced subspace. To arrive at the reduced Hamiltonian, we apply the $\widetilde{P}$ projection
on both the input and output states of the Hamiltonian operator. As the state is represented
in the $\widetilde{B}$ basis, one can either convert the vector to $\Theta$,
project into $\widetilde{\mathcal{H}}$, apply $H$, project the resulting
state again into $\widetilde{\mathcal{H}}$, and transform it to $\widetilde{B}$,
or, equivalently, first apply the projection and then the base transformation.
These two alternatives correspond to the following two expressions for $H_{1}$:
\begin{equation}\label{eq:H1_def}
H_{1}:=\widetilde{G}^{\dagger}\left(\widetilde{P}H\widetilde{P}\right)\widetilde{B}=\widetilde{P}_{\widetilde{B}B}\left(G^{\dagger}HB\right)\widetilde{P}_{B\widetilde{B}}.
\end{equation}
and
\begin{equation}
H_{1}=\widetilde{G}^{\dagger}H\widetilde{B}=\left(\widetilde{B}^{\dagger}\widetilde{B}\right)^{-1}\left(\widetilde{B}^{\dagger}H\widetilde{B}\right).
\label{eq:H_tilde_B_tilde_B}\end{equation}

The TISE and TDSE then take the form
\begin{equation}
E\psi_{\widetilde{B}}=H_{1}\psi_{\widetilde{B}}\label{eq:TISE_red}
\end{equation}
\begin{equation}
\partial_{t}\psi_{\widetilde{B}}=-\frac{i}{\hbar}H_{1}\psi_{\widetilde{B}}.\label{eq:TDSE_red}
\end{equation}
\emph{Eq. \ref{eq:H_tilde_B_tilde_B}, \ref{eq:TISE_red} and \ref{eq:TDSE_red} will be the key working equations for the remainder of this manuscript.}

The formulation using only $\widetilde{B}$ matrices (r.h.s. of \ref{eq:H_tilde_B_tilde_B})
is appealing, as $\widetilde{B}=BR$ is the naturally reduced matrix
in $\widetilde{P}$.

Note that $H_{1}$ is similar to a Hermitian matrix, in that the product of matrices to the left and right of $H$ in eq. \ref{eq:H_tilde_B_tilde_B} produce the identity, $\left(\left(\widetilde{B}^{\dagger}\widetilde{B}\right)^{-1}\widetilde{B}^{\dagger}\right)\left(\widetilde{B}\right)={\IdentOp}$. Therefore the eigenvalues are real and the evolution is unitary. Moreover, combining eq. \ref{eq:H_tilde_B_tilde_B} and \ref{eq:TISE_red} the TISE may be reformulated as a generalized eigenvalue equation with Hermitian matrices,
\begin{equation}
\left(\widetilde{B}^{\dagger}\widetilde{B}\right)\lambda\psi_{\widetilde{B}}=\left(\widetilde{B}^{\dagger}H\widetilde{B}\right)\psi_{B}, \label{eq:Scrod_eig_red}
\end{equation}
thus avoiding the matrix inversion required to compute $\widetilde{S}$.
This form is solvable with an iterative eigensolver, such as Arnoldi.
Unfortunately, in this form it is not possible to restrict the solver's
operations to the more efficient matrix-vector  (i.e. operator-state) multiplications.

While accurate and mathematically rigorous, computing the $H_{1}$
form in eq. \ref{eq:H_tilde_B_tilde_B} is time consuming. Most significantly,
calculation of elements of the reduced Hamiltonian $\widetilde{B}^{\dagger}H\widetilde{B}$
can be laborious, particularly in the multi-dimensional case, as the
$B$ functions are non-localized. Fortunately, there are symmetry
considerations and numerical techniques which can accelerate this
computation by several orders of magnitude, discussed at length in the following section. Alternatively, using eqs. \ref{eq:H_tilde_B_tilde_B}, \ref{eq:Overlap-matrices} and \ref{eq:tilde_S_and_invS}, one may rewrite $H_{1}$ as
\begin{equation}
H_{1}=\widetilde{S}\left(R^{\dagger}S^{-1}G^{\dagger}HGS^{-1}R\right).\label{eq:H1_on_the_fly}
\end{equation}
Eq. \ref{eq:H1_on_the_fly} may be quicker to compute then \ref{eq:H1_def}: due to the locality of $G$ and
the fact that elements of $H$ are usually functions of either position or
momentum (but rarely both), the vast majority of $G^{\dagger}HG$
elements are vanishingly small. Note that $G$ and $S^{-1}$ decompose dimensionally.
Indeed, one would still have to calculate $\widetilde{S}$, which requires
the inversion of $\widetilde{B}^{\dagger}\widetilde{B}$, but this may be
accelerated if one is able to store the matrix. If we then use an iterative algorithm
for solving the TISE (e.g. Arnoldi) and the TDSE (e.g. Taylor propagator), one may use only matrix-vector
type operations to further accelerate the process. Similar considerations have been noted in  \cite{TuckerCarrington}.

There are several possible approximations to \ref{eq:TDSE_red}. These are discussed in detail in \cite{The-Math-Paper}.

\section{\label{sec:Algorithms}Practical considerations and Algorithms}

The task of computing the reduced Hamiltonian, $H_1$,

\begin{equation}
H_{1}=\widetilde{G}^\dagger H \widetilde{B} = \left(\widetilde{B}^{\dagger}\widetilde{B}\right)^{-1}\widetilde{B}^{\dagger}H\widetilde{B},
\label{eq:H_1}\end{equation}

(eq. \ref{eq:H1_def}), for multi-dimensional systems can be computationally expensive.

As the reduced basis is adapted dynamically throughout the computation, $H_{1}$ must be updated on-the-fly. It is therefore vital that this computation be performed as efficiently as possible. In Section \ref{sub:Computing-Hamiltonian-elements} we discuss several techniques to accelerate this computation, each of which provides over an order-of-magnitude improvement over the naive approach. Algorithms for solving the TISE and TDSE, which utilize the ideas presented in Section \ref{sub:Computing-Hamiltonian-elements}, are discussed in Sections \ref{sub:Eigenmode-algorithm} and \ref{sub:Dynamics},
respectively.

\subsection{\label{sub:Computing-Hamiltonian-elements}Evaluating the reduced Hamiltonian}

Computing multi-dimensional Hamiltonians that include the factor $\widetilde{B}^{\dagger}H\widetilde{B}$, as per eq. \ref{eq:H_tilde_B_tilde_B} may be significantly accelerated by decomposing the Hamiltonians into a sum of products of one dimensional Hamiltonians. This is discussed in Section \ref{sub:Decomposition-POTFIT}. Further speed-up may be achieved by an iterative update scheme for the inverse matrix $\left(\widetilde{B}^{\dagger}\widetilde{B}\right)^{-1}$ of eq. \ref{eq:H_tilde_B_tilde_B}, described in Section \ref{sub:Updating S}. By utilizing the symmetries in the Hamiltonian, and symmetries inherent to Gabor functions $G$ and $B$, better than an order-of-magnitude reduction in computational costs is possible, as detailed in Section \ref{sub:Utilizing-symmetries-in-H}.

On top of these, the process of computing new Hamiltonian matrix elements lends itself to parallelization - both in a multi-core and multi-machine configuration, which further accelerates the process.

The cells in phase space will be labeled by their $2\times d\times N_{\textrm{p}}$ phase space coordinates.

\subsubsection{\label{sub:Decomposition-POTFIT}Decomposition of the Hamiltonian
into a sum of one dimensional products (POTFIT)}

The conversion of the Hamiltonian from the pseudo-spectral representation
(as approximated by sampling) to the reduced $\widetilde{B}$ basis can
be broken down into two parts: first, the construction of the Hamiltonian
matrix,
and second the computation of $\widetilde{S}=\left(\widetilde{B}^{\dagger}\widetilde{B}\right)^{-1}$.
In this section we shall concentrate on the former, and in the next
section on the latter.

In general, $H$ may be written as a sum of terms, some of which depend
on a single Degree of Freedom (DoF) such as linear momentum$\frac{\hbar^{2}k_{x}^{2}}{2m}$,
and some of which depend on multiple DoF such as the Coulomb potential

$\frac{q_{1}q_{2}}{4\pi\epsilon_{0}}\left(\left(x_{1}-x_{2}\right)^{2}+\left(y_{1}-y_{2}\right)^{2}+\left(z_{1}-z_{2}\right)^{2}\right)^{-\frac{1}{2}}$.

Single DoF terms are easy to compute. As they
are functions of either $x$ or $k$ (but not both), the relevant
Hamiltonian term is diagonal in either the pseudospectral or spectral
bases, and therefore the multiplication $b_{j}^{\dagger}Hb_{k}$
for such terms requires only vector-vector operations.
However, to convert the terms depending on multiple DoF to $\widetilde{B}$ would require multi-dimensional
integration, which may be extremely costly.
We would therefore like to replace the multi-DoF terms with a sum of products of
one-dimensional terms.

Most often, the non-decomposable parts of the Hamiltonian are the
particle-particle interaction terms. We therefore seek a decomposition of
the multi-dimensional interaction tensor as

\begin{equation}
V\left(x_{1},x_{2}\ldots x_{N}\right)=\sum_{j_{1}=1}^{m_{1}}\ldots\sum_{j_{N}=1}^{m_{n}}c_{j_{1}\ldots j_{N}}\left(t\right)V_{j_{1}}^{\left(1\right)}\left(x_{1}\right)\otimes\ldots\otimes V_{j_{N}}^{\left(N\right)}\left(x_{N}\right)
\end{equation}

with the single-particle potential vectors, $V^{\left(k\right)}$,
normalized and $c$ representing the magnitude of each term. The
problem of finding an optimal decomposition (with minimal
possible error for any number of terms) is an open problem, and outside
the scope of this work. For our purposes, we shall utilize the POTFIT
algorithm \cite{POTFIT-1,POTFIT-2}, which is optimal for two degrees of freedom
and applicable generally.

\subsubsection{\label{sub:Updating S}Updating $\widetilde{S}$}

The reduced basis is updated dynamically as the static and  dynamic
algorithms progress (Sections \ref{sub:Eigenmode-algorithm} and \ref{sub:Dynamics-algorithm}).
Since the number of cells added or removed from the overlap matrix $\widetilde{S}$
in each iteration is relatively low compared to the number
of elements already in $\widetilde{S}$, one may save significantly on
computational effort by updating $\widetilde{S}$ to account for the change,
rather than recomputing it from scratch.

When adding $\widetilde{M}\ll\widetilde{N}$ rows and columns to the existing $\widetilde{N}\times\widetilde{N}$
matrix $\widetilde{S}$, we shall reorder matrix indexes such that the
new elements are to the right and below the existing elements.
If $\widetilde{S}^{-1}$ is available from a previous iteration, the expanded matrix's inverse, which we shall term
$\widetilde{Z}^{-1}$, may be computed by \cite{Matrix-Inv-by-Blocks}

\begin{eqnarray}
\widetilde{Z}_{\left(\widetilde{N}+\widetilde{M}\right)\times\left(\widetilde{N}+\widetilde{M}\right)}^{-1} & = & \left(\begin{array}{cc}
\widetilde{S}_{\widetilde{N}\times\widetilde{N}} & C_{\widetilde{N}\times\widetilde{M}}\\
C_{\widetilde{M}\times\widetilde{N}}^{\dagger} & D_{\widetilde{M}\times\widetilde{M}}
\end{array}\right)^{-1}=\left(\begin{array}{cc}
\widetilde{S}^{-1}+F^{\left(2\right)}C^{\dagger}\widetilde{S}^{-1} & -F^{\left(2\right)}\\
F^{\left(1\right)}C^{\dagger}\widetilde{S}^{-1} & F^{\left(1\right)}
\end{array}\right)\label{eq:mat_inv_add_1}
\end{eqnarray}

where

\begin{eqnarray}
F_{\widetilde{M}\times\widetilde{M}}^{\left(1\right)} & := & \left(D-C^{\dagger}\widetilde{S}^{-1}C\right)^{-1},\label{eq:mat_inv_add_2}\\
F_{\widetilde{N}\times\widetilde{M}}^{\left(2\right)} & := & \widetilde{S}^{-1}CF^{\left(1\right)}.\label{eq:mat_inv_add_3}
\end{eqnarray}

Note that at no point do we invert a matrix of dimension $\widetilde{N}$,
and that all matrix multiplications include at least one dimension
$\widetilde{M}\ll\widetilde{N}$.

The inverse operation - removing $\widetilde{M}$ rows and columns (which,
after appropriate reordering, may be assumed to be on the right and
below the elements remaining), divides the larger, previously known,
$\widetilde{Z}^{-1}$ into four blocks
\begin{equation}
\widetilde{Z}_{\left(\widetilde{N}+\widetilde{M}\right)\times\left(\widetilde{N}+\widetilde{M}\right)}^{-1}=\left(\begin{array}{cc}
W_{S,\widetilde{N}\times\widetilde{N}} & W_{C,\widetilde{N}\times\widetilde{M}}\\
W_{C,\widetilde{M}\times\widetilde{N}}^{\dagger} & W_{D,\widetilde{M}\times\widetilde{M}}
\end{array}\right).
\end{equation}

From eq. \ref{eq:mat_inv_add_1}, \ref{eq:mat_inv_add_2} and \ref{eq:mat_inv_add_3}
we derive
\begin{equation}
\widetilde{S}^{-1}=W_{S}+W_{C}W_{D}^{-1}W_{C}^{\dagger}
\end{equation}

In typical dynamics computations, there are two or three orders of
magnitude separating $\widetilde{M}$ and $\widetilde{N}$. The above scheme
provides a better than order-of-magnitude acceleration in updating
the inverse vs. re-computing it from scratch.

\subsubsection{\label{sub:Utilizing-symmetries-in-H}Utilizing symmetries in the
Hamiltonian and the Gabor basis}

The $\mathcal{G}$ basis is Gabor (eq. \ref{eq:PvN-basis}). If we
choose the von Neumann lattice points to be a subset of the FG sample
points ($x$) and spectral frequencies ($k$), then the discrete $G$
and $B$ matrices are also Gabor, and therefore highly symmetric.
Moreover, there are multiple symmetries within each of the Hamiltonian matrix
elements, resulting in multiple cells of the reduced Hamiltonian
having identical values. This may be utilized to accelerate computation
of $\widetilde{B}^{\dagger}H\widetilde{B}$ by more than an order of magnitude.

Let us consider a single term in a Hamiltonian which has been decomposed
into a sum of products of one-dimensional terms.
In practice, as most such terms in the Hamiltonian stem from a POTFIT
decomposition of multi-dimensional interaction potentials (see Section
 \ref{sub:Decomposition-POTFIT}), we will focus on $V\left(x\right)$-type
terms.

For purposes of the following discussion, the columns of the $\widetilde{B}$
matrix shall be enumerated not by a single number, but by their phase space
lattice coordinates. Similarly, we shall view $\widetilde{B}^{\dagger}H\widetilde{B}$
as a 4-dimensional phase space tensor (see Section \ref{sub:Computing-Hamiltonian-elements}).
Gabor basis functions have a momentum dependence of the form $e^{ik_{\beta}\mod_{L}\left(x-x_{\alpha}\right)}$.
Therefore

\begin{equation}
\begin{split}
\left(\widetilde{B}^{\dagger}V\widetilde{B}\right)&_{x_{\alpha}k_{\beta}x_{\gamma}k_{\delta}} = \sum_{x_{m}\in F.G.}B_{x_{\alpha}k_{\beta}}^{*}\left(x_{m}\right)V\left(x_{m}\right)B_{x_{\gamma}k_{\delta}}\left(x_{m}\right) \\
 & = \quad e^{-i\left(k_{\delta}-k_{\beta}\right)\mod_{L}\left(x_{\alpha}-x_{\gamma}\right)}\sum_{x_{m}\in F.G.} B_{x_{\alpha}k_{0}}^{*}\left(x_{m}\right)V\left(x_{m}\right)B_{x_{\gamma}k_{0}}\left(x_{m}\right)e^{i\left(k_{\delta}-k_{\beta}\right)x_{m}}.
\end{split}
\end{equation}

Note the dependence is only on $k_{\delta}-k_{\beta}$, and not on
the individual momenta. Moreover, if there are symmetries in $V$,
which there often are (e.g. particle exchange symmetry in the Coulomb
potential), these too can be taken into account. Finally, $\widetilde{B}^{\dagger}H\widetilde{B}$
is Hermitian, providing an additional symmetry.

In principle, treating the kinetic term $T$ is similar, with the roles of $x$ and $p$ exchanged.
Therefore, in many instances the value of the $\widetilde{B}^{\dagger}H\widetilde{B}$ cell
has already been computed previously.

We incorporate these symmetries into the algorithm by creating a canonical
phase space coordinate for each $\widetilde{B}^{\dagger}H\widetilde{B}$ cell with
the same value. For example we address $\widetilde{B}^{\dagger}H\widetilde{B}$ cells
by $\left(\widetilde{B}^{\dagger}H\widetilde{B}\right)_{x_{j_{1}},p_{k_{1}}-p_{k_{2}},x_{j_{2}},0}$,
converting the 4D coordinate into a 1D linear coordinate for easy
comparison. The above methodology can be extended directly to the
multi-dimensional case.

\subsection{\label{sub:Eigenmode-algorithm}Statics: The eigenmode algorithm}

Given a potential, the initial task is to find its ground-state, and
possibly higher excited states. Since we do not have a priori information on the eigenstate,
we use an iterative process, starting with an initial guess at the local minima of the potential.
We then iteratively expand the subspace, until the designated eigenmodes of the reduced Hamiltonian
fall off to below the prescribed amplitude at the boundary of the subspace.
Note that no classical considerations are used.

The TISE solving algorithm is as follows:

\medskip{}

\texttt{10 Define the initial reduced basis as the cells at the $x$-s
of the potential's local minima, with $k=0$.}

\texttt{20 Compute eigenmodes for the current reduced basis.}

\texttt{30 If everywhere on the boundary of the reduced basis, all
eigenmode cell amplitudes are below the specified accuracy threshold,
stop. If not, continue.}

\texttt{40 Remove all phase space cells where the amplitude is below
the wavefunction accuracy cutoff.}

\texttt{50 Expand the reduced basis to all neighboring cells (i.e. all
cells at or below some distance $r$ from the current reduced basis),
and compute the new entries of the now expanded, reduced Hamiltonian
matrix.}

\texttt{60 Go to 20.}

\medskip{}

A few additional comments:
\begin{itemize}
\item The radius  of the reduced basis defines a $2\times d\times N_{\textrm{p}}$-dimensional
ball associated with the local neighborhood of each phase space cell. In step $\texttt{50}$ of the algorithm, we  center this ball at each cell in the reduced basis which is above the amplitude cutoff, to determine the new, expanded, reduced basis. It has been our experience that a radius of $\sqrt{2}$ ($+\epsilon$ to account for numerical noise), produces the most stable results.
\item There is no need to fully diagonalize the reduced Hamiltonian in step
$\texttt{20}$, as often we are only interested in a few low-lying
states. Therefore, one may resort to the much-faster Arnoldi/Lanczos
methods.
\item In multi-dimensional systems, the unoptimized algorithm spends the majority
of its time computing elements of the reduced Hamiltonian (step
$\texttt{50}$). See above for methods to accelerate this step.
\item The algorithm above proves to be extremely stable, including cases
of multiple local minima. However, no proof of convergence is currently
known.
\end{itemize}
Finally, note that for the eigenvalue problem, there is no need to
explicitly invert $\widetilde{B}^{\dagger}\widetilde{B}$. Instead, one may
solve the generalized eigenvalue equation $\widetilde{B}^{\dagger}H\widetilde{B}\psi_{\widetilde{B}}=E\left(\widetilde{B}^{\dagger}\widetilde{B}\right)\psi_{\widetilde{B}}$.
as discussed in Section \ref{sub:The-TDSE-for-the-reduced_state}.

\subsection{\label{sub:Dynamics}Dynamics}

\subsubsection{\label{sub:Dynamics-algorithm}Dynamics algorithm}

A wavefunction's evolution in time is continuous, and always proceeds
only to neighboring cells in the von Neumann lattice. Given an initial
state in a reduced subspace, one can modify the reduced subspace containing
the wavefunction on-the-fly, adjusting it as the state evolves. This
may be done by monitoring the wavefunction amplitude at the phase space
boundary of the reduced space. If cells rise above the specified accuracy
threshold, for example at the ``bow'' of a traveling wave-packet,
the reduced basis is expanded in that region of the boundary. Conversely,
as the amplitude falls below the threshold at the wavepacket's ``stern'',
vectors are removed from the reduced basis.

Tunneling presents no problem to the method. The onset of tunneling is apparent as motion
through the barrier. The accuracy cutoff is low enough to avoid discarding
the exponentially decreasing wavefunction amplitude inside the classically
forbidden area within the reduced subspace.

The algorithm is presented below. The choice of propagator for step
$\texttt{20}$ is discussed in the following section.

\medskip{}

\texttt{10 Set the initial state (usually the ground state) and initial
reduced basis.}

\texttt{20 Propagate the reduced state with the reduced Hamiltonian
for $\Delta t$.}

\texttt{30 If everywhere on the boundary of the reduced state the
amplitude is below the specified accuracy threshold, go to 20. If
not, continue.}

\texttt{40 Remove all phase space cells where the amplitude is below
the accuracy threshold.}

\texttt{50 Expand the reduced basis to all neighboring cells (i.e. all
cells at or below some distance $r$ from the current reduced basis): Compute
the additional elements of the now expanded reduced Hamiltonian.}

\texttt{60 Go to 20.}

\medskip{}

\subsubsection{\label{sub:Propagators}Time propagation}

Until now we have focused on efficient methods for evaluating $H_1$ (eq. \ref{eq:H_1}). We now
turn to evaluating the TDSE, eq. \ref{eq:TDSE_red}, i.e. choosing an efficient propagator
given the unique features of PvB. Specifically, if one is propagating a state in the reduced
PvB basis, there is a need to modify the reduced basis as the wavefunction
changes, and therefore to modify the $\widetilde{S}$ and $\widetilde{B}^{\dagger}H\widetilde{B}$
matrices (see eqs. \ref{eq:H_tilde_B_tilde_B} and \ref{eq:TDSE_red}).
This in turn imposes limits of the duration of the allowed time step,
turning the task into a series of short-time propagations.
This subject is discussed in depth in Section \ref{sub:Dynamical-timestep-for}.
In the case of time-varying control fields, where the Hamiltonian
is composed of a drift and control Hamiltonians, $H\left(t\right)=H_{\textrm{d}}+H_{\textrm{c}}\left(t\right)$,
the short-time step requirement above suggests that we should approximate the
time-dependent Hamiltonian by a series of piecewise constant Hamiltonians.
This approximation is valid given that by the very definition of the
reduced subspace, we are removing components of the state whose amplitude
is below some pre-defined amplitude (e.g. $10^{-6}$). In conclusion,
we are dealing with short time step propagation, with a pre-defined
limit on required propagation accuracy.

Many propagators are possible in such a scenario, including split-operator,
Lanczos-type methods (e.g. Short Iterative Lanczos and Short Iterative
Arnoldi), Inverse-Free Lanczos and direct Taylor series expansion.
Given our need for a short-time propagation method, we have found the Taylor series
expansion to be both the most efficient and simplest to implement,
with clear and direct control over accuracy. Below is a detailed discussion
of the methods examined.

\subsubsection*{Split Operator }

By prediagonalizing the control Hamiltonian, $H_{\textrm{c}}=W_{1}D_{1}W_{1}^{-1}$
and pre-exponentiating the drift, $U_{0}=\exp\left(\tau H_{\textrm{d}}\right)$,
we may write (omitting $-\frac{i}{\hbar}$ everywhere)

\begin{equation}
\exp\left(\tau\left(H_{0}+u_{1}H_{1}\right)\right)=W_{1}e^{u_{1}D_{1}}W_{1}^{-1}U_{0}W_{1}e^{u_{1}D_{1}}W_{1}^{-1}+\mathcal{O}\left(\tau^{3}\right).
\end{equation}

\begin{figure}
\noindent \begin{centering}
\includegraphics[scale=0.6511]{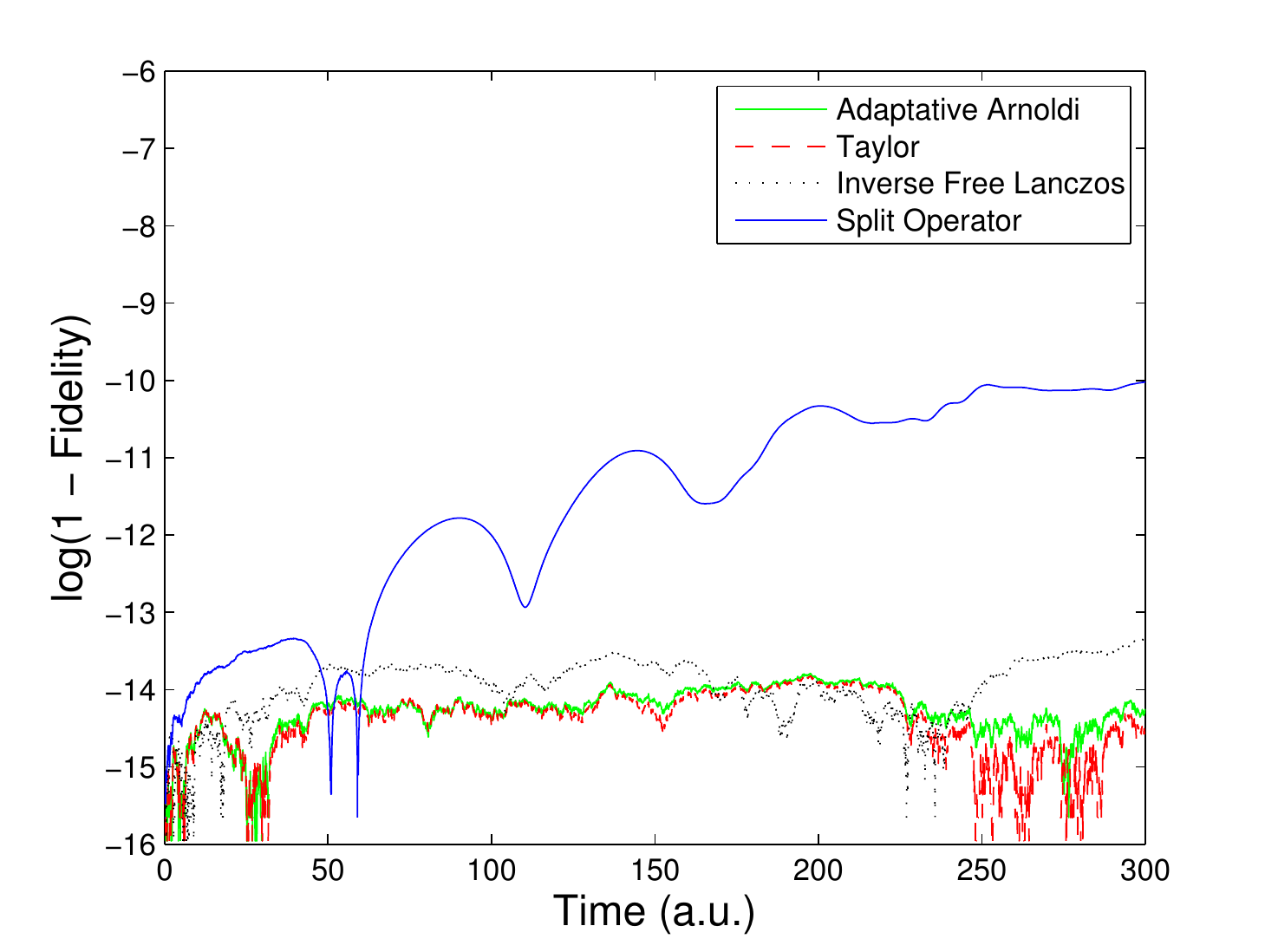}
\par\end{centering}

\noindent \begin{centering}
\protect\caption{\label{fig:Propagator-accuracy-and-speed}Comparison of propagator
accuracy. Note that Adaptive Arnoldi, Taylor and Inverse-free Lanczos
all achieve excellent accuracy, while the split-operator suffers
from a non-trivial drift, stemming from the $\mathcal{O}\left(\tau^{3}\right)$
error inherent to the method. Given these equivalent accuracies, the
simplicity of the Taylor propagator, and the ability to easily control
its accuracy (by dynamically varying both the number
of terms in the series and the time step), make it the method of choice
for PvB propagation.}

\par\end{centering}

\noindent \centering{}
\end{figure}

Once the preliminary work is done, one may propagate the state, accounting for a time-dependent $u\left(t\right)$
without resorting to additional matrix exponentiations or diagonalizations. However,
within the reduced PvB framework, we must redo the prediagonalization step
at each basis change, drastically reducing its efficiency. Moreover,
the split operator suffers from the accumulation of the $\mathcal{O}\left(\tau^{3}\right)$
error (see fig. \ref{fig:Propagator-accuracy-and-speed}), which is
expensive to mitigate by reducing $\tau$. Therefore the method is
both inaccurate and slow for our purposes.

\subsubsection*{Lanczos-type methods }

Krylov methods, such as Short Iterative Lanczos or Short Iterative
Arnoldi, rely on the construction, at each time step, of a small subset
of the Hilbert space called the Krylov space. The Krylov space is obtained by
applying the Hamiltonian to the state iteratively several times. Then, one exponentiates the small Krylov
Hamiltonian. The dimension of the Krylov space should be adapted dynamically
to maintain a constant accuracy.

Unfortunately, it is not possible to directly use a Lanczos-type algorithm
with PvB, as the effective Hamiltonian to propagate, $H_{1}=\widetilde{G}^{\dagger}H\widetilde{B}$,
is not Hermitian, but only similar to a Hermitian matrix. Two resolutions
are possible: first, one may apply the Short Iterative Arnoldi (SIA)
method, which allows non Hermitian Hamiltonians. We have found the
SIA method to be comparable in accuracy and performance to the Taylor
series expansion (below). However, it is significantly more complex
to implement, and provides no discernible advantage.

One may avoid the somewhat costly inversion of $\widetilde{S}$ by looking
at the problem as a generalized eigenvalue problem with a Hermitian
$\widetilde{B}^{\dagger}H\widetilde{B}$ Hamiltonian, and utilizing the Inverse-Free
Lanczos (IFL) method  \cite{IF-Lanczos}. IFL replaces the inversion
in the Lanczos algorithm by a Cholesky decomposition (which is slightly
faster) and the upper triangular Hamiltonian of the Arnoldi algorithm
with a tri-diagonal Krylov Hamiltonian (which presents a linear system
requiring solution - a non trivial task). However, as the full inversion
of $\widetilde{S}$ is not required at every modification of the reduced
basis (Section \ref{sub:Updating S}), IFL's advantage is significantly
reduced.

\subsubsection*{Taylor series expansion}

The simplest propagation method is the expansion of the exponential in a Taylor
series, which may be rephrased as a recursion series. Denoting the $k$-th derivative of $\psi$ w.r.t time by $\psi^{\left(k\right)}$, we have
\begin{eqnarray}
\psi^{\left(0\right)} & = & \psi\left(t\right) \nonumber \\
\psi^{\left(k\right)} & = & \frac{\tau}{k}H\psi^{\left(k-1\right)} \nonumber \\
\psi^{\left(t+\tau\right)} & = & \sum_{k=0}^{\infty}\psi^{\left(k\right)}.
\end{eqnarray}

This method entails no matrix-matrix operations, but simply vector-matrix multiplications.
Moreover, for reduced PvB dynamics, we may formulate the recurrence
relation as
\begin{equation}
\psi^{\left(k\right)}=\frac{\tau}{k}\widetilde{S}\left(\widetilde{H}\psi^{\left(k-1\right)}\right),
\end{equation}

thus avoiding the costly $\widetilde{S}\widetilde{H}$ matrix-matrix multiplication,
and replacing it with two matrix-vector operations, which are significantly
quicker. Finally, the order (number to terms) to which the Taylor
series is computed is determined dynamically, by requiring that $\left\Vert \psi^{\left(k\right)}\right\Vert \le\epsilon\ll1$.
Should the series extend beyond some reasonable limit (e.g. $30$
terms), the propagator signals the caller that the time step $\tau$
needs to be decreased (see Section \ref{sub:Dynamical-timestep-for})
This allows highly accuracy propagation with minimal computational
resources, and is our method of choice for PvB.

\subsubsection{\label{sub:Dynamical-timestep-for}Multi-factor determination of
dynamic time step for PvB evolutions}

In order to achieve efficient propagation of the state in the PvB
scheme, one needs to take three issues into account.
\begin{enumerate}
\item PvB dynamics will occasionally indicate that the reduced PvB basis needs
to be updated (expanded or reduced), which will necessitate
a change in the size of the state vector and computation of new elements
in $\widetilde{B}^{\dagger}H\widetilde{B}$ before propagation can proceed. See
Sections \ref{sub:Dynamics-algorithm} and \ref{sub:Computing-Hamiltonian-elements}.
\item Most propagator methods, such as the Taylor series expansion, are
accurate and efficient only up to a certain propagation duration.
\item Approximating the control signal by a piecewise constant function, the propagation step
may not exceed the bound dictated by the control signal discretization,
which in turn is determined by its derivative (see below).
\end{enumerate}
We therefore dynamically manage the propagation time step as follows:
\begin{enumerate}
\item A maximal time step is defined based on the gradient of the control
signal. See discussion below.
\item If a cell that was added to the boundary during the previous time step
is found to be above the wavefunction amplitude cutoff, $\zeta$, after the propagation
step, the last propagation is cancelled, the time step is reduced by a factor of 2 and then the propagation is redone. Conversely, if after
a small number of time steps there is no expansion of the basis, then
the time step is increased gradually (by $20\%$), subject to the
limit imposed by the discretization of the control pulse, above.
\item If the propagator concludes that the time step is too large for it to perform
the propagation accurately and efficiently (e.g. the number of elements
in the Taylor series reached $30$), it signals the integration loop,
which then discards this propagation and redoes it with a smaller
integration interval (usually half the original time step).
\end{enumerate}
Finally, as mentioned above, the maximal time step limit may be derived from the discretization
of the control signal. Taking as an example a $P$ (momentum) kick
control field,

\begin{equation}
\left|\psi\right\rangle =\exp\left[i\mathcal{T}\int_{0}^{T}(H_{\textrm{d}}+u(t)H_{\textrm{c}})dt\right]\left|\psi_{0}\right\rangle \,,
\end{equation}
with $\mathcal{T}$ indicating time-ordering, $T\ll1$ and $H_{\textrm{c}}=P=i\partial_{x}$.
A first order approximation gives:
\begin{equation}
\left|\psi\right\rangle \approx\int_{0}^{T}u(t)\partial_{x}\psi(x,t)dt\left|\psi_{0}\right\rangle \,,
\end{equation}
 which suggests the constraint:
\begin{equation}
-U_{m}KT\leq\int_{0}^{T}u(t)\partial_{x}\psi(x,t)dt\leq U_{m}KT\,,
\end{equation}
 with $K$ the maximum spatial frequency of $\psi$ and $U_{m}=\max|u|$.
As $T\ll1$ we can approximate $U_{m}\approx T\partial_{t}u(t)$ which
finally produces the condition:
\begin{equation}
\left|\psi_{max}\right\rangle -\left|\psi_{min}\right\rangle =2KT^{2}\partial_{t}u(t)\left|\psi_{0}\right\rangle \leq\zeta\,,
\end{equation}
 or a weaker but more useful condition:
\begin{equation}
T\leq\sqrt{\frac{\zeta}{2K\max_{t}\left(\partial_{t}u(t)\right)}}\,.
\end{equation}
For example, in the case of the one dimensional model of helium detailed in \ref{sub:Helium-double-ionization}, for $\zeta=10^{-4}$, $K\approx0.2$a.u. and $\max_{t}\left(\partial_{t}u(t)\right)\approx2.5$, we obtain $T\approx0.01$ a.u.

\section{\label{sec:Examples}Examples}

\subsection{\label{sub:Double-well-potential}Double-well potential}

\begin{figure}
\noindent \begin{centering}
\includegraphics[scale=0.1034]{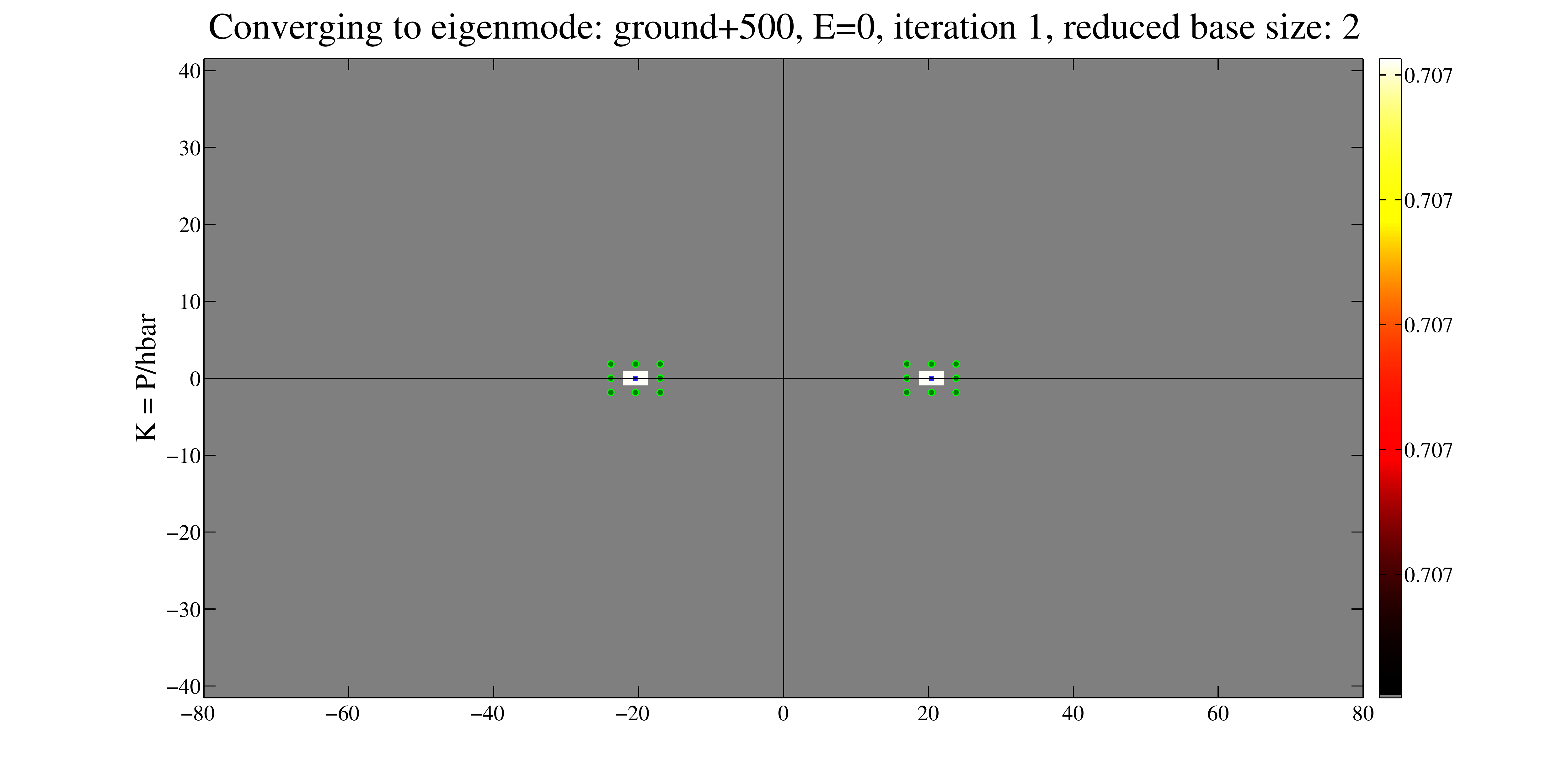}\includegraphics[scale=0.1034]{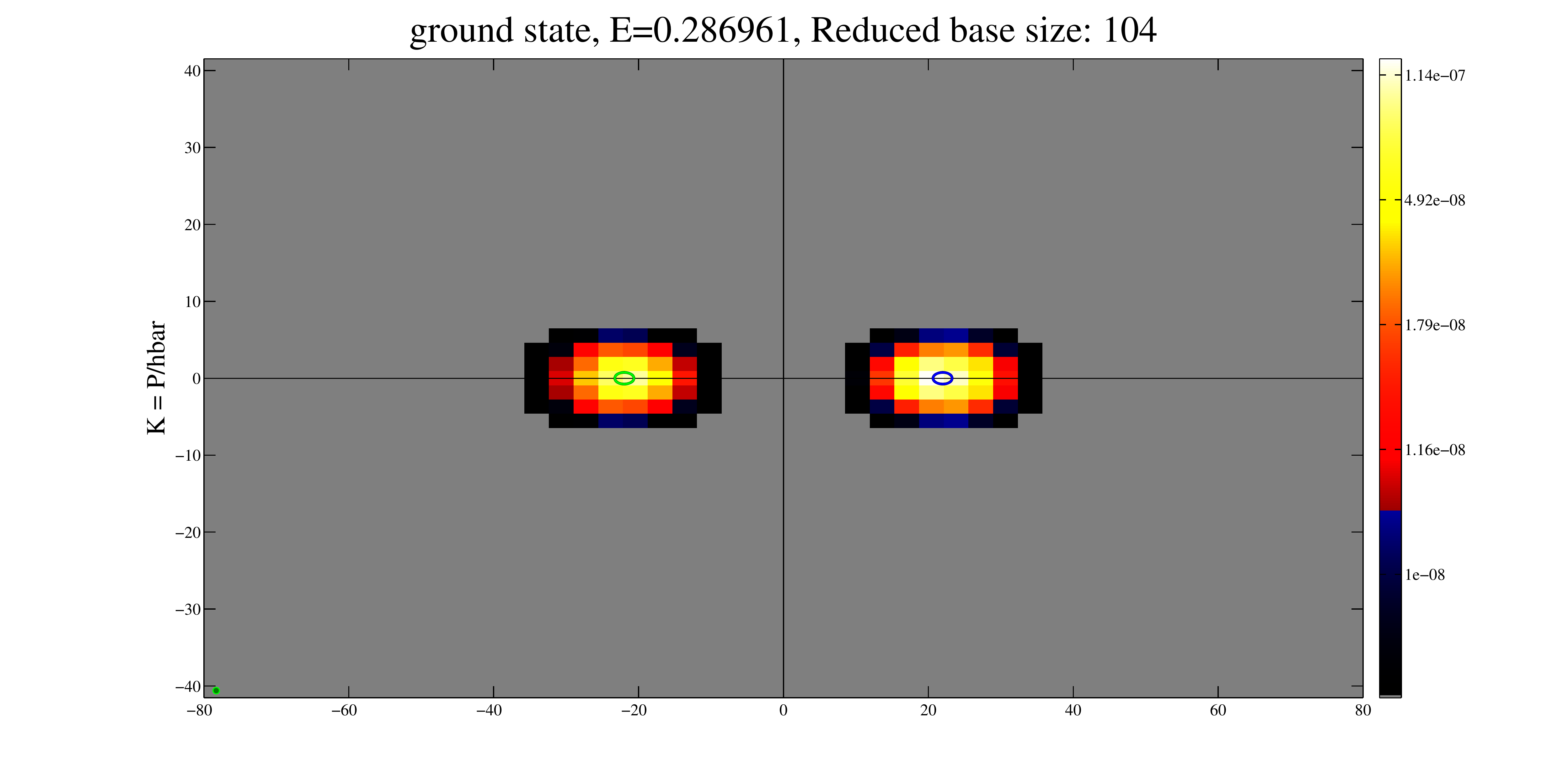}\\
\includegraphics[scale=0.1034]{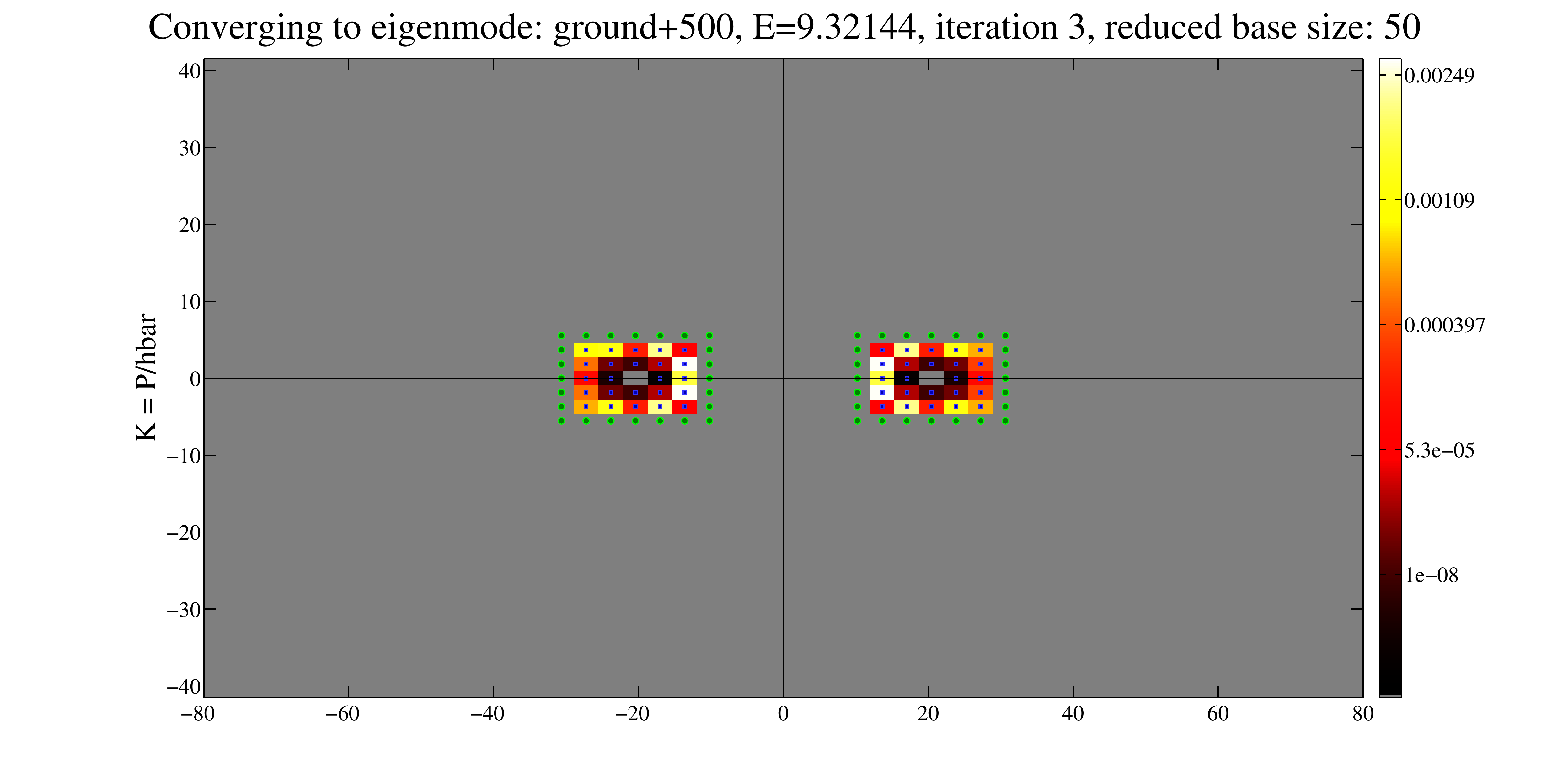}\includegraphics[scale=0.1034]{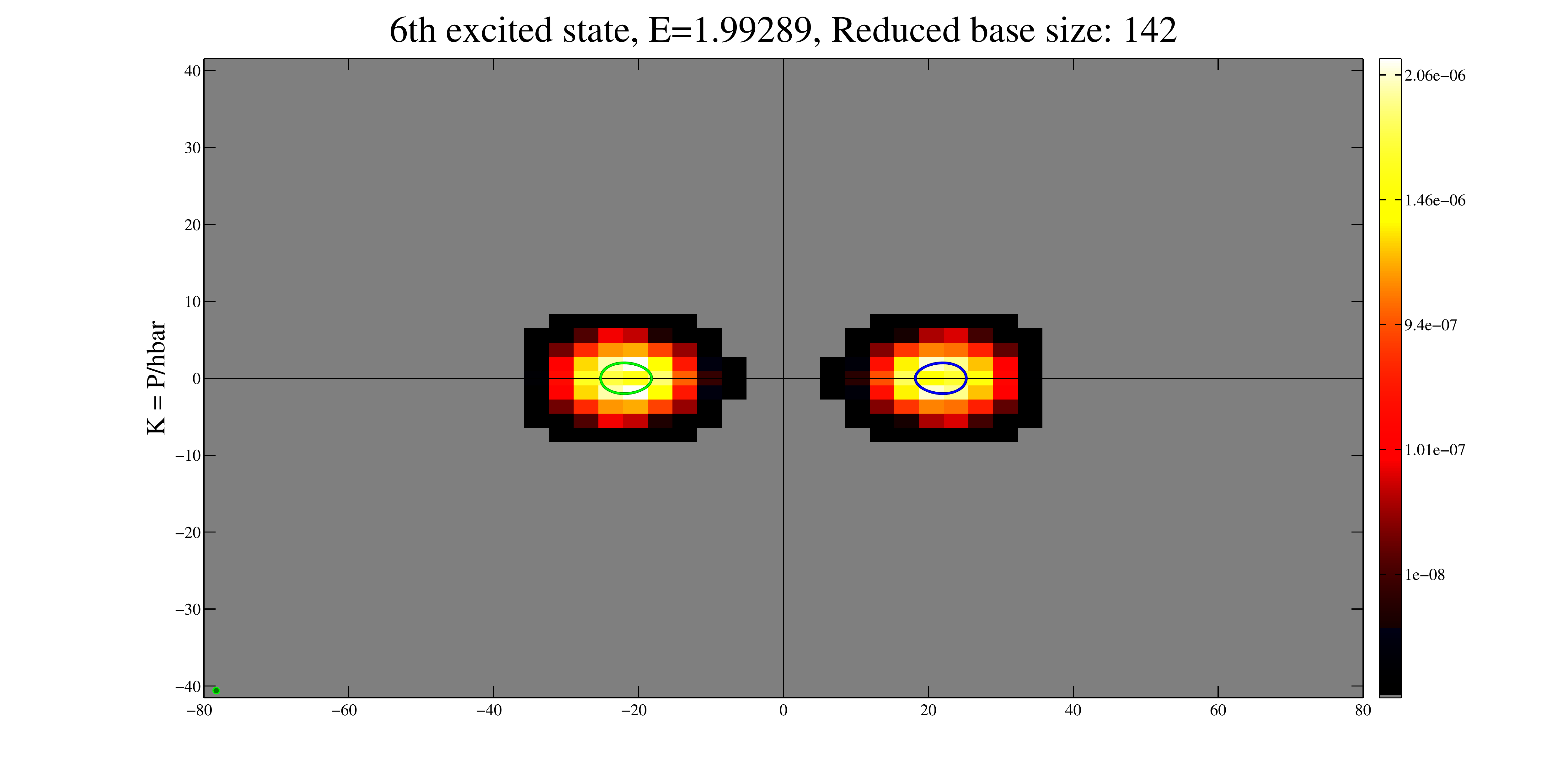}\\
\includegraphics[scale=0.1034]{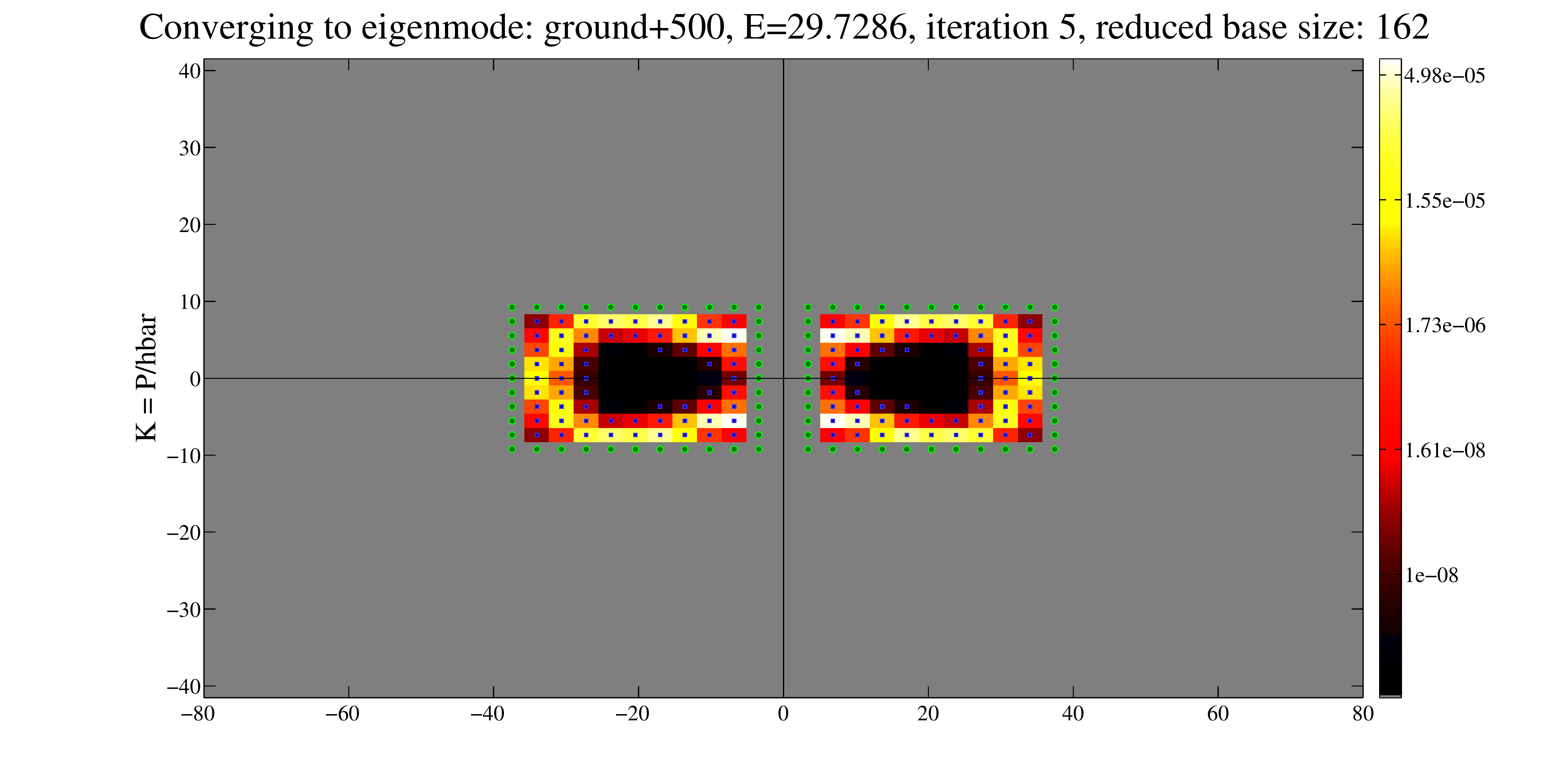}\includegraphics[scale=0.1034]{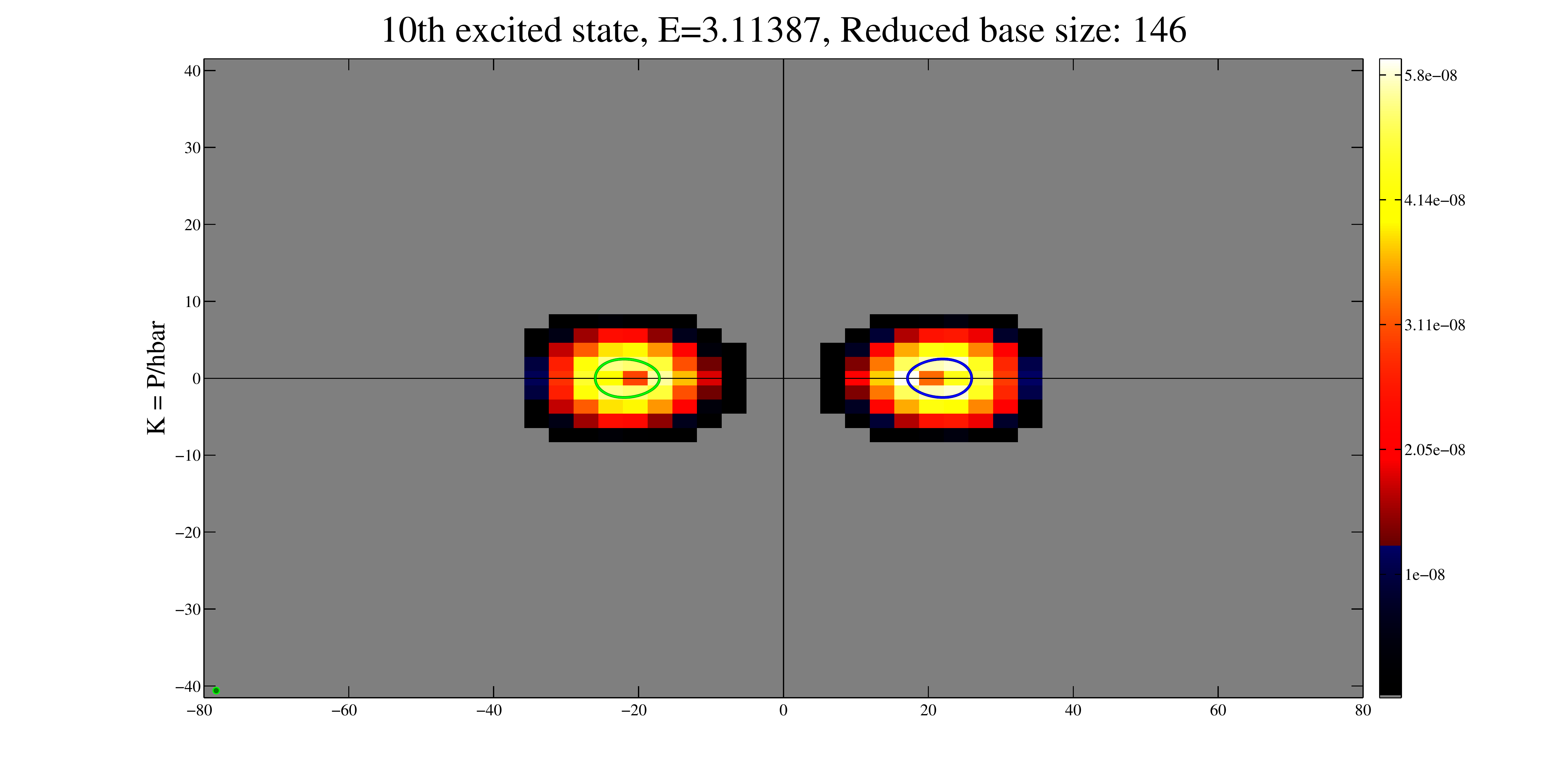}\\
\includegraphics[scale=0.1034]{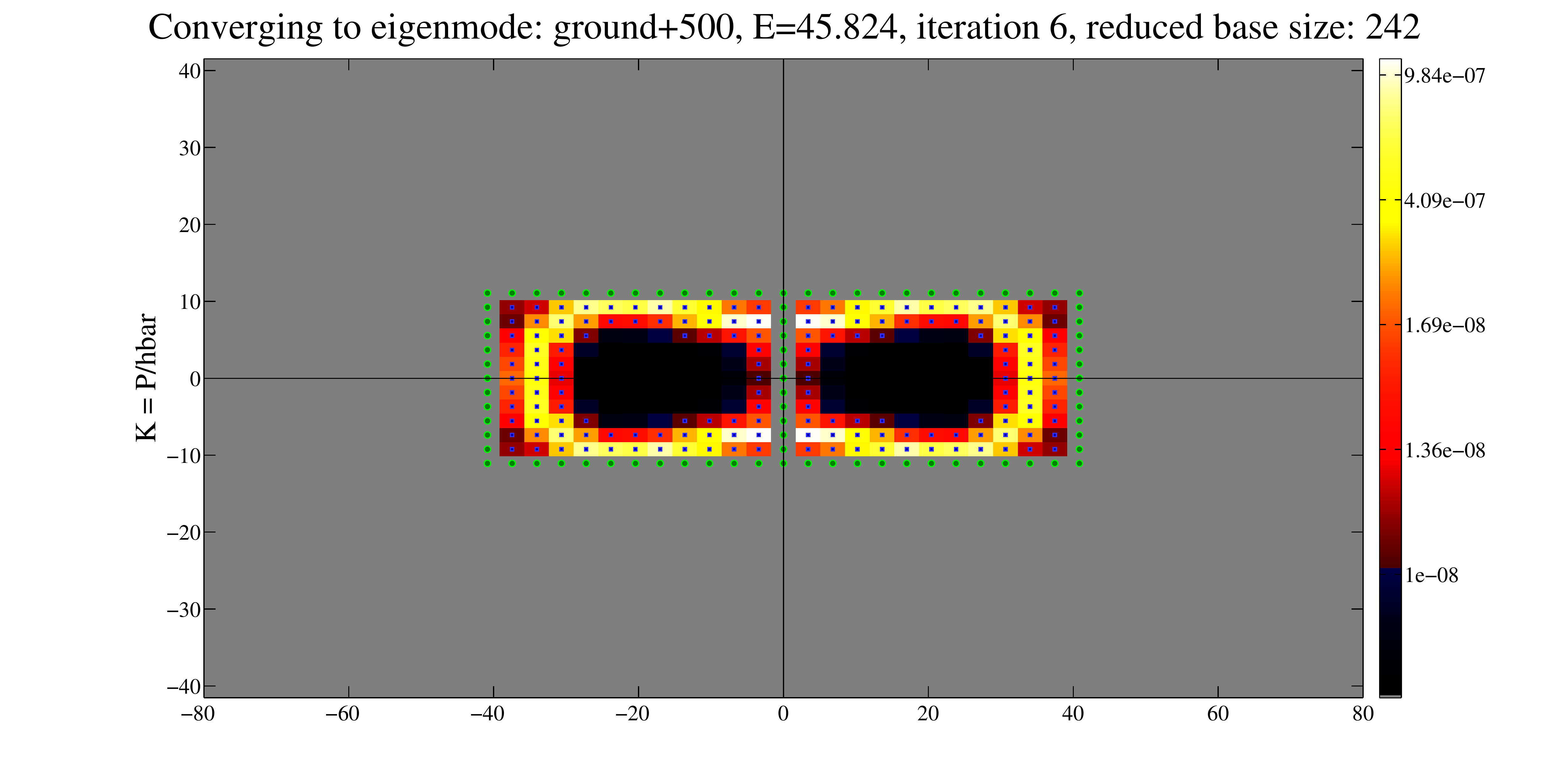}\includegraphics[scale=0.1034]{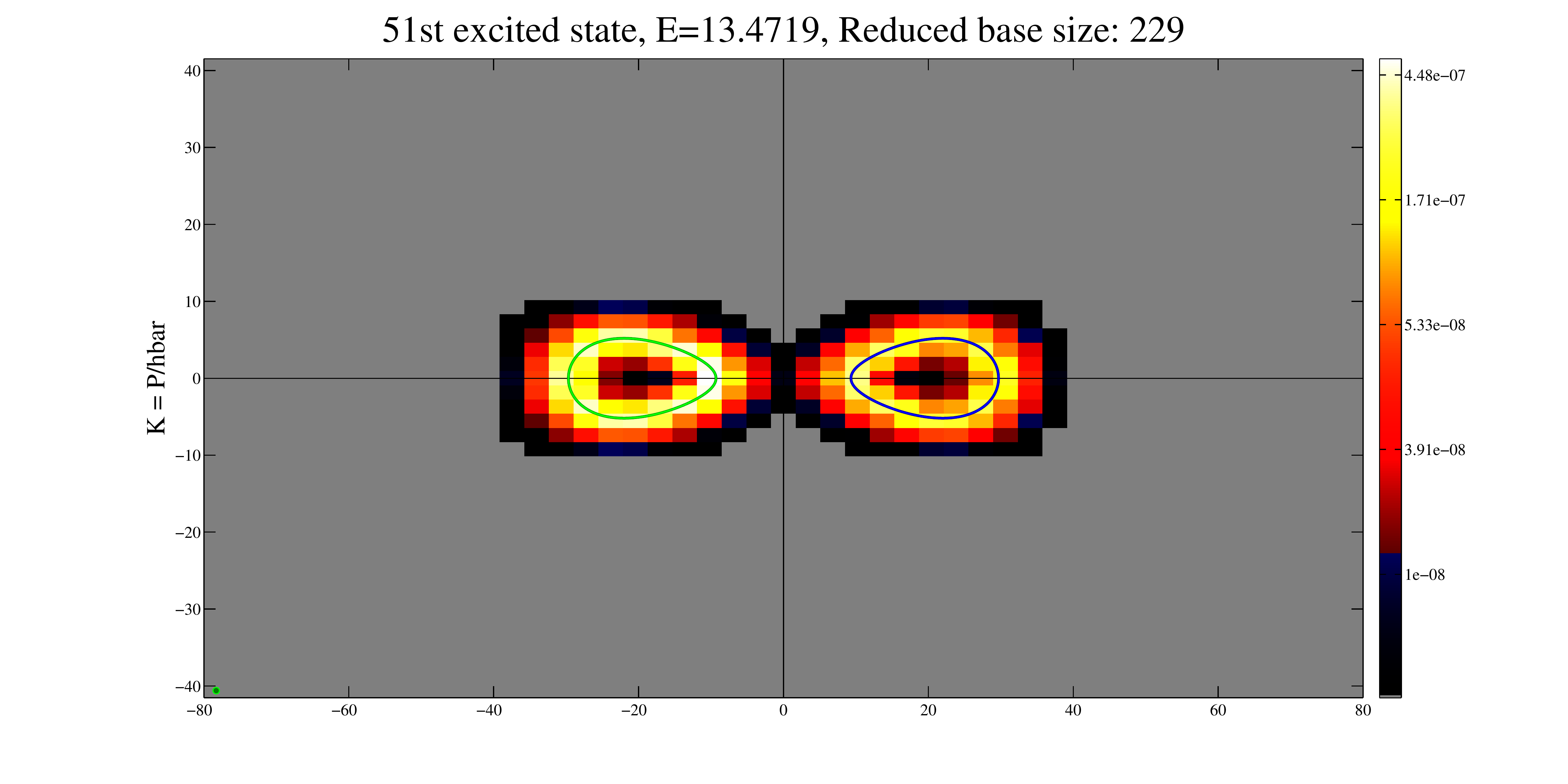}\\
\includegraphics[scale=0.1034]{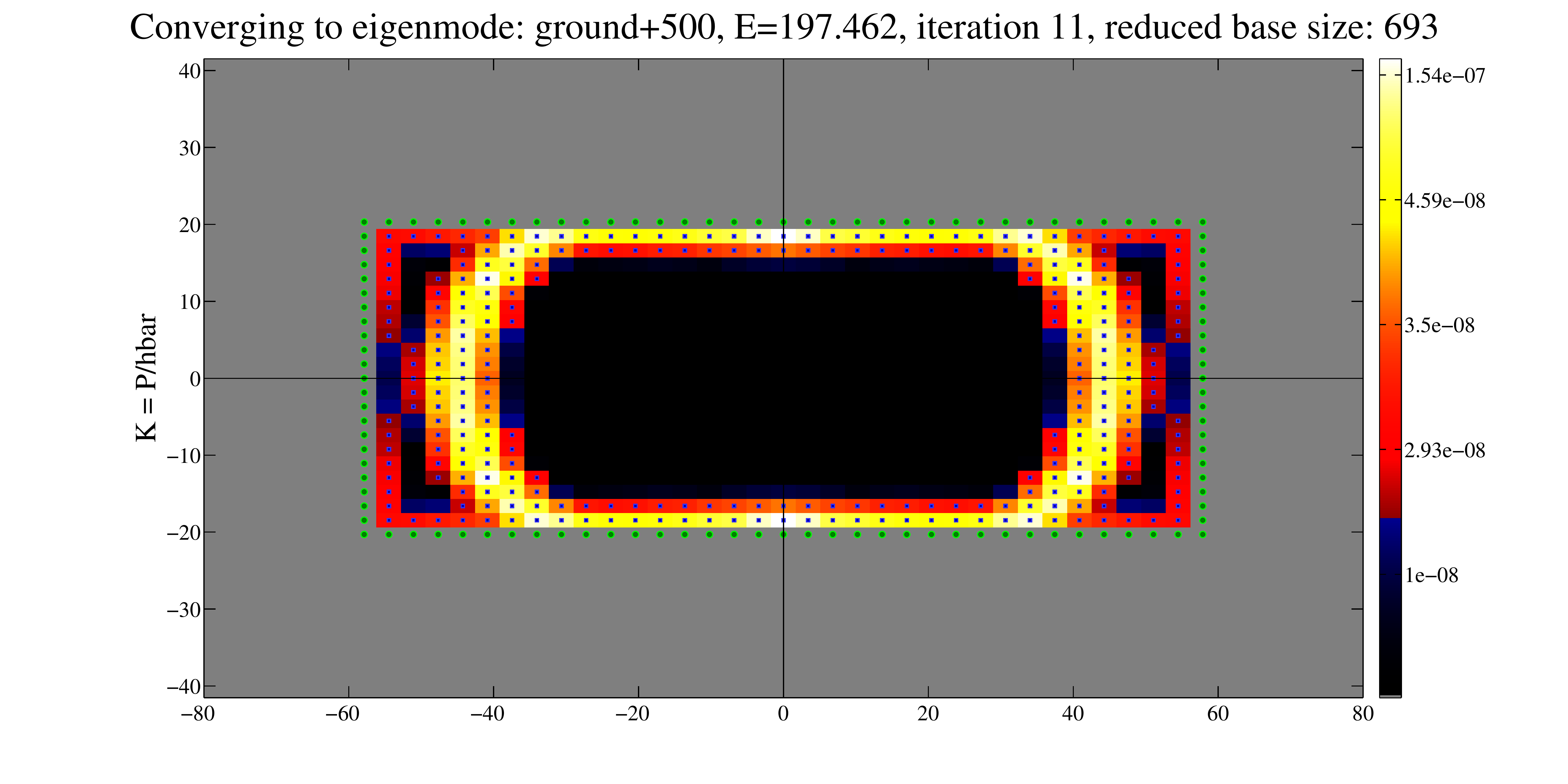}\includegraphics[scale=0.1034]{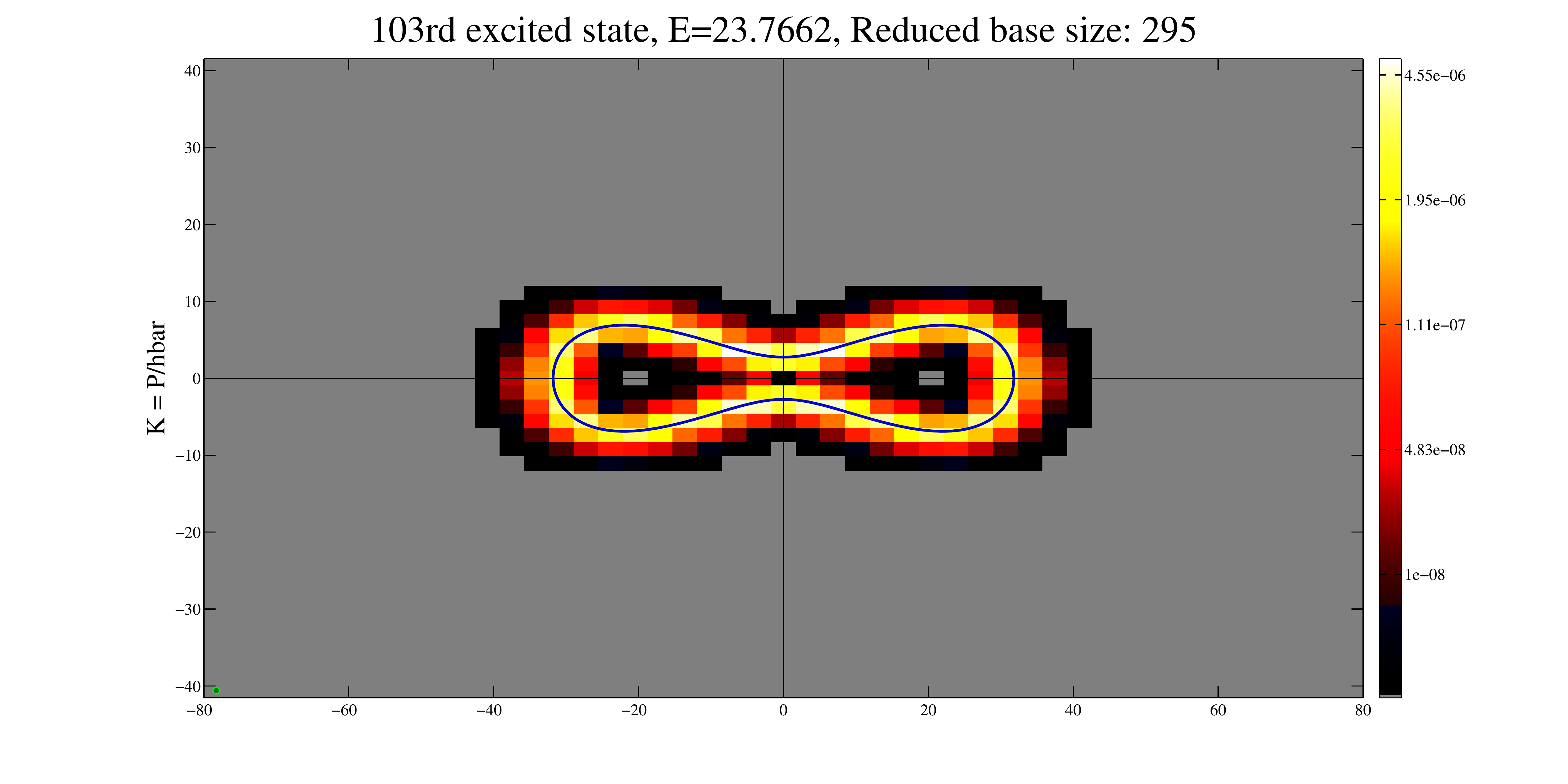}\\
\includegraphics[scale=0.1034]{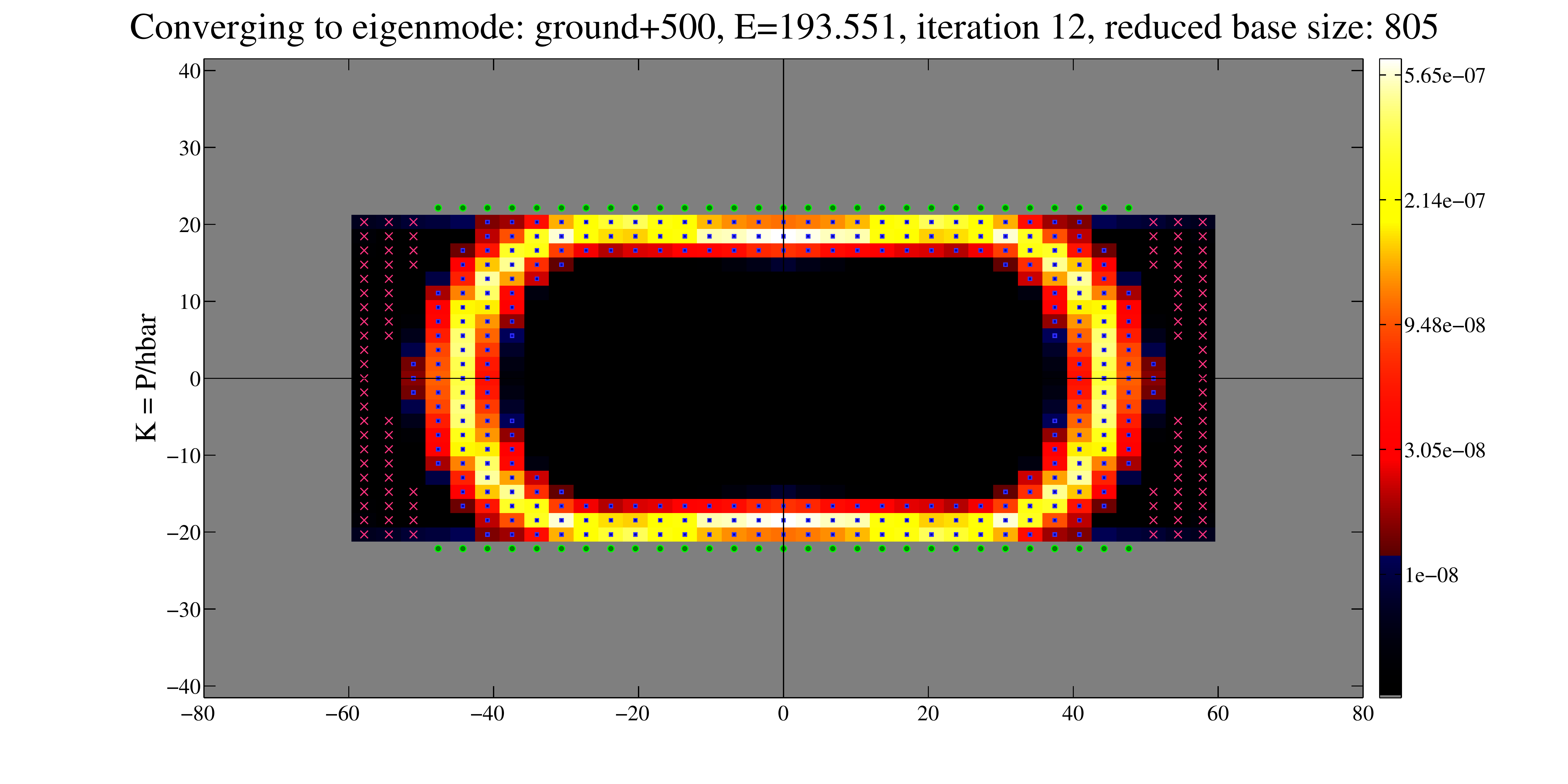}\includegraphics[scale=0.1034]{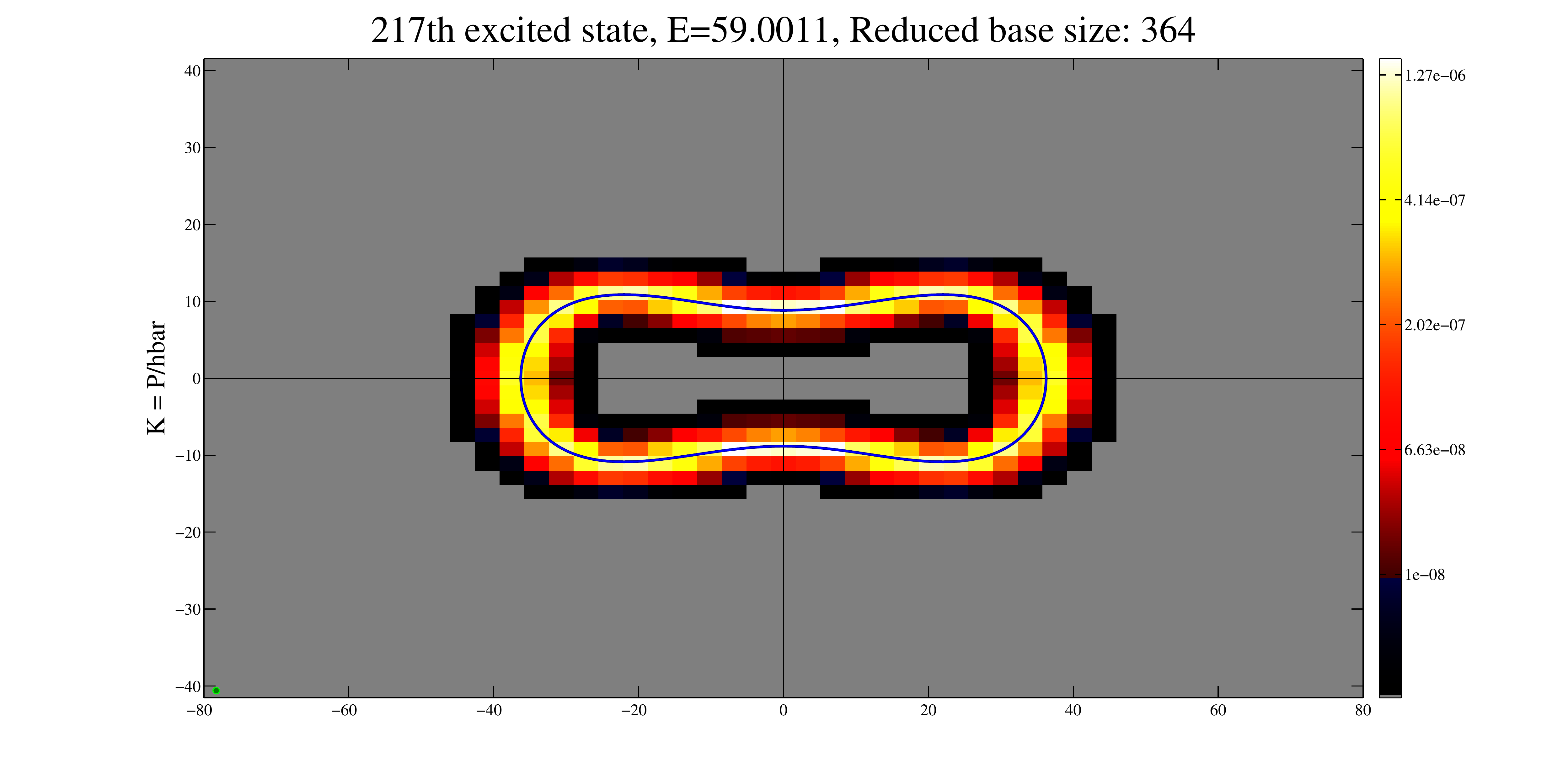}\\
\includegraphics[scale=0.1034]{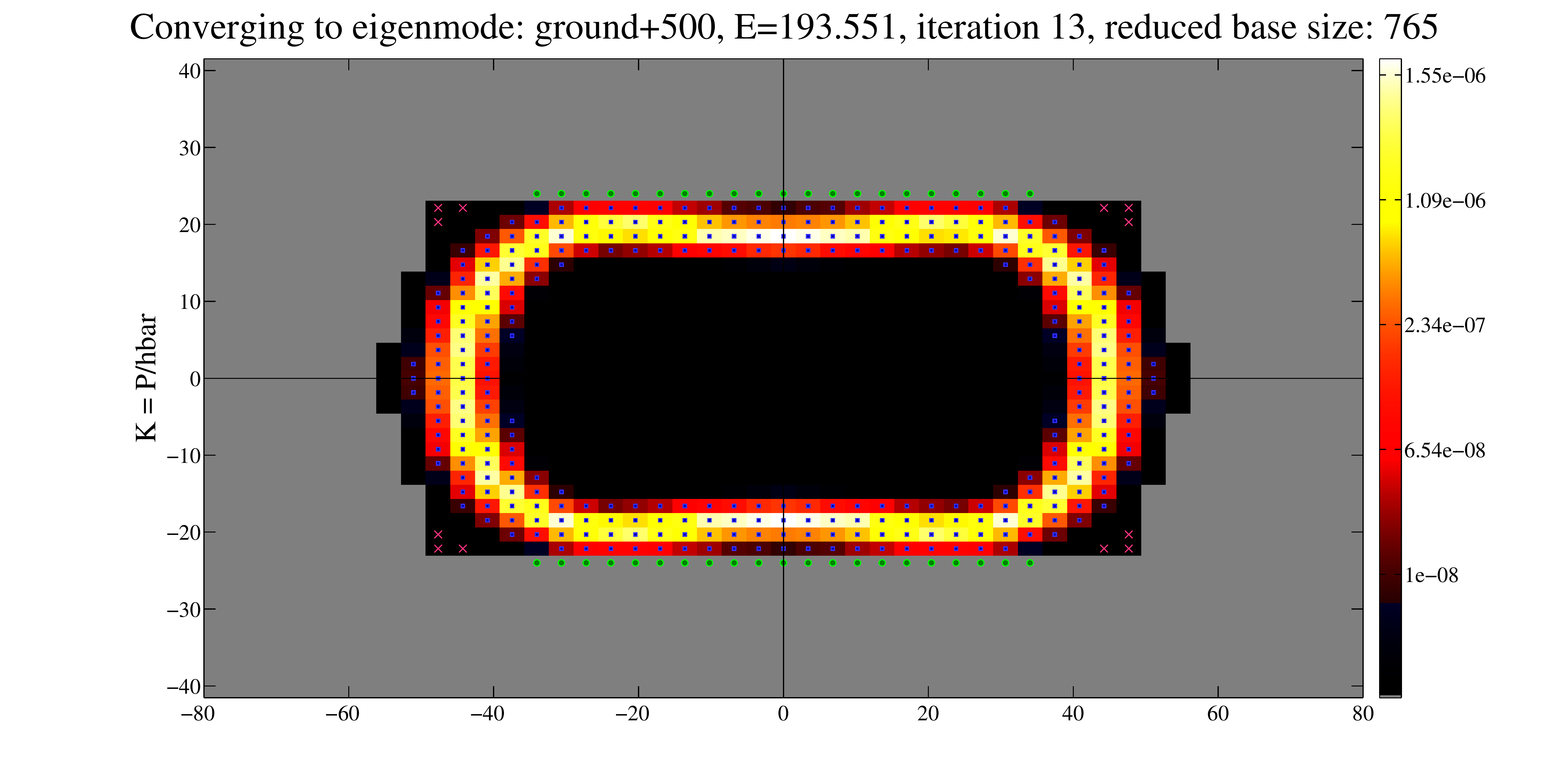}\includegraphics[scale=0.1034]{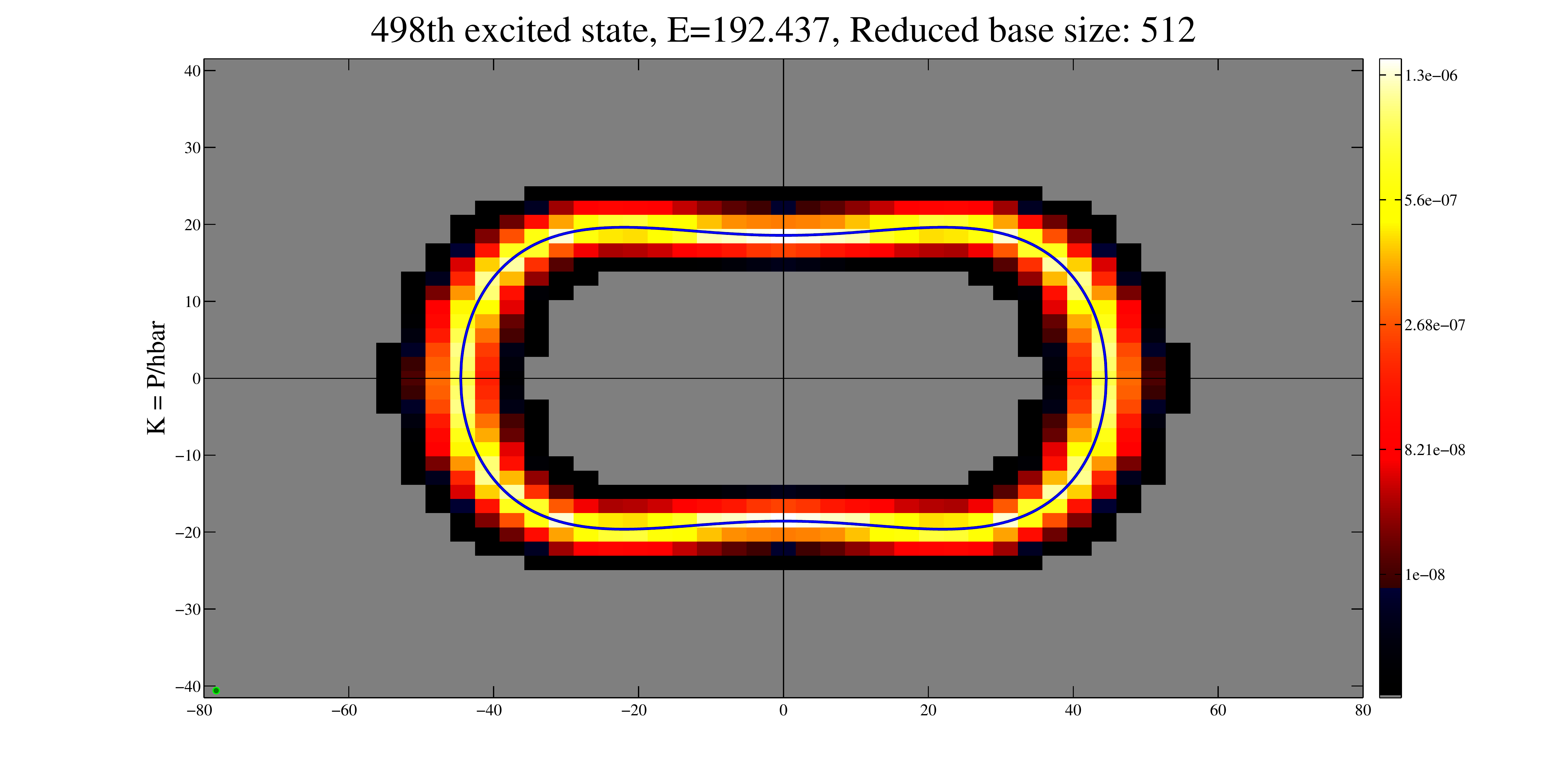}
\par\end{centering}

\protect\caption{\label{fig:Double-well-potential}Convergence process (left) and eigenmodes
(right) for the symmetric double-well potential. Left column contains
$7$ of $13$ steps in the iterative solution of the TISE, as detailed
in Section \ref{sub:Eigenmode-algorithm}. On the right we see $7$
of the $500$ eigenmodes computed, with the blue and green lines denoting
the classical energy contours at the eigenmode energy. Gray area denotes
phase space which is outside the reduced subspace. Color of each pixel
denotes the absolute value amplitude of the overlap of the von Neumann
grid Gaussian centered at the pixel and the eigenmode in question,
i.e. $\protect\braket{g_{\bar{x}_{k},\bar{p}_{k}}^{\textrm{mod}}}{\psi}$.}

\end{figure}
We now demonstrate the PvB method by applying it to the TISE for a
double-well potential. Specifically
\begin{equation}
V\left(x\right)=m\omega^{2}b\left(\left(\frac{x}{d}\right)^{4}-2\left(\frac{x}{d}\right)^{2}+1\right),
\end{equation}

where $V\left(0\right)=b$ and the local minima are $V\left(\pm d\right)=0$,
and where $m=1$, $\omega=1$, $b=20$ and $d=22$. The Hamiltonian
is $H=\frac{p^{2}}{2m}+V\left(x\right)$. The lowest 500 eigenmodes
are computed using the algorithm in Section \ref{sub:Eigenmode-algorithm}.

Figure \ref{fig:Double-well-potential} shows the eigenmode convergence
process (left column) and the resultant eigenmodes (right column).
The iterative solution to the TISE begins with the reduced phase space
defined as the local minima of the potential. With each iteration
the phase space can both expand and contract, as needed. From high-amplitude
cells it is iteratively expanded to neighboring cells, until the eigenmode
amplitude at the boundary falls below some predefined threshold. Cells
with red x-s will be removed from the reduced space before the next
iteration, cells with green circles will be added to the reduced phase space
in the next iteration, and cells with blue dots will remain untouched.
The large low-amplitude (dark) area in the middle of the state is
not pruned as it contains lower-lying states, which are computed concurrently, but
not plotted.

The right column depicts some of the resulting eigenmodes. Note the transition from
a solution similar to two adjacent harmonic oscillators at lower energies, to a single well solution
at higher energies. Also note the correspondence of the classical energy contours (green and blue lines)
and the quantum mechanically occupied phase space.

While the eigenmodes plotted are highly pixelated, intuitively suggesting a lower accuracy solution,
this is not the case. The eigenmodes of the potential in question
are well localized in phase space, and are accurately represented
within the Fourier grid chosen. The pixelization is just a visual feature of the method,
reflecting the size of the phase space basis functions.
The only approximation comes from the explicit cutoff of the wavefunction amplitude which defines the reduced subspace (here $\zeta=10^{-6}$).

\subsection{\label{sub:Helium-double-ionization}Helium double ionization}

To demonstrate the power of the PvB methodology, we turn to an example
where the phase space hyper-rectangle required to bound the dynamics (the Fourier grid) is far
greater than the phase space volume which the state occupies. In such
a scenario, the reduced phase space is many orders of magnitude smaller
than the unreduced Hilbert space. Specifically, we shall look at the
problem of multi-electron ionization.

The topic of multi-electron dynamics and its control have been of great
interest in recent years. This is largely due to progress made in the field of attosecond
physics, specifically the availability of extreme-ultraviolet
(XUV) attosecond pulses  \cite{XUV-atto-via-HHG} created via high
harmonic generation with a high intensity near-infrared (NIR) femtosecond
laser pulse. As the natural timescale of bound electrons is in the
attosecond range  \cite{atto-timescale}, this opens up the possibility
of probing and manipulating both internal molecular dynamics and
atomic ionization processes. Subsequent studies utilized simultaneous
NIR and XUV pulses to control the dynamics  \cite{XUV-NIR-experiment-1,XUV-NIR-experiment-2}.

This problem remains a significant numerical challenge, even for the
smallest multi-electron system, i.e. the helium atom. Strong NIR pulses
populate high angular momentum levels, making the use of hyperspherical based
methods inefficient, while the traveling wavepackets of the ionized
electron requires a very large spatial grid. In such scenarios, the
PvB method demonstrates its strength, requiring resources proportional
only to the occupied area of phase space.

In this work we shall only briefly present the problem and show some
sample results of TDSE simulations. A detailed study of both
concerted and sequential multi-electron ionization of helium using
PvB will be published by the authors separately  \cite{PvB-MCTDH-comparison-paper}.

Our benchmark system, the 1D helium model, consists of two electrons,
each with a single degree of freedom, interacting with each other
and a central (nuclear) potential. We use a regularized form of the
Coulomb potential, $\frac{1}{\sqrt{r^{2}+a_{0}^{2}}}$, where the
regularizer, $a_{0}=0.739707902$, is such that the ground-state energy
of the model matches the experimentally measured binding energy of
helium, $2.903385\,\textrm{ amu}$  \cite{NIST-fund-const,NIST-spectroscopic-data}.
The Hamiltonian is therefore:
\begin{equation}
H=\frac{1}{4\pi\epsilon_{0}}q_{e}Q\left(\frac{1}{\sqrt{x_{1}^{2}+a_{0}^{2}}}+\frac{1}{\sqrt{x_{2}^{2}+a_{0}^{2}}}\right)+\frac{1}{4\pi\epsilon_{0}}\frac{q_{e}^{2}}{\sqrt{\left(x_{1}-x_{2}\right)^{2}+a_{0}^{2}}}+\frac{1}{2m_{e}}\left(p_{1}^{2}+p_{2}^{2}\right)\label{eq:He1D Hamiltonian}
\end{equation}
with $Q=-2q_{e}$ and $q_{e}$ being the electron charge.

We solve the TDSE in eq. \ref{eq:TDSE_red}, \ref{eq:H_tilde_B_tilde_B},
by transforming the Hamiltonian to the $\widetilde{B}$
basis via $\widetilde{B}^{\dagger}\left(T+V\right)\widetilde{B}$. In the
case of multi-electron dynamics, the electron-electron interaction
term, $\frac{1}{\sqrt{\left(x_{1}-x_{2}\right)^{2}+a_{0}^{2}}}$,
is not dimensionally decomposable, and therefore extremely expensive
to integrate numerically. We represent it as a sum-of-products of one dimensional potentials
(although in this particular case a very large number of elements
are required to achieve an accurate representation ). POTFIT was used
to perform this decomposition, see Section \ref{sub:Decomposition-POTFIT}.

\begin{figure}
\noindent \begin{centering}
\includegraphics[scale=0.8402]{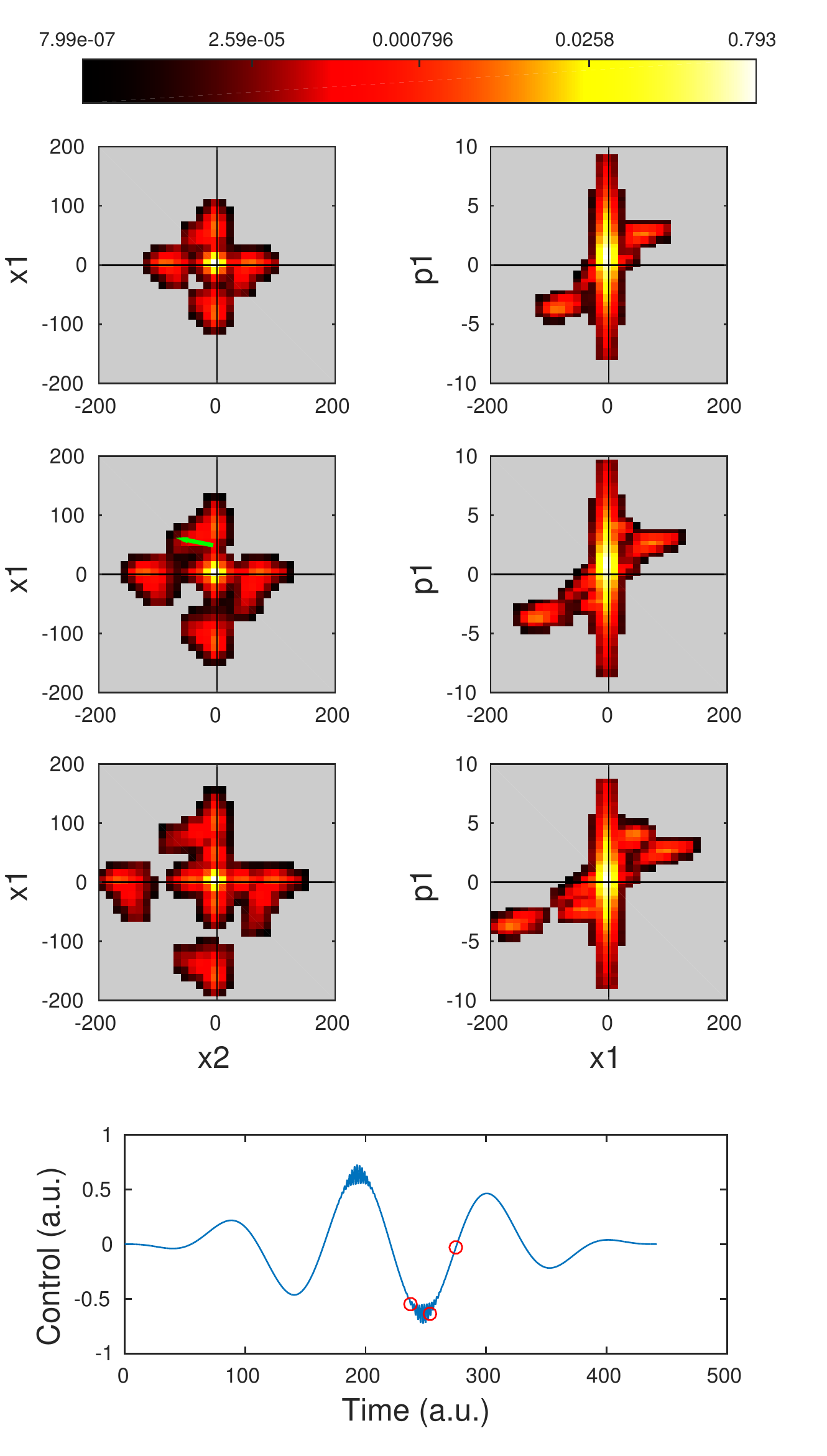}
\par\end{centering}

\noindent \centering{}\protect\caption{\label{fig:H1D_TDSE_progression}Snapshots of the two electron wavefunction
after the second XUV pulse, sorted by increasing times from top to
bottom. The first column shows the $x_{1};x_{2}$ correlations. The
second column shows the one dimensional phase space projection of
the first electron. The times corresponding to the three snapshots
are pointed out by red circles on the control pulse in the bottom
frame. The green arrow in the middle-left plot points to a concerted
ionization contribution.}
\end{figure}

The system is driven by a combination of a NIR pulse and two XUV pulses
(see lower panel of fig. \ref{fig:H1D_TDSE_progression}). The NIR
pulse is taken to have a sine envelop in order to have exactly zero
derivatives at the beginning and at the end: $u_{\textrm{\textrm{NIR}}}=A_{\textrm{\textrm{NIR}}}\sin(2\pi t/T_{\textrm{\textrm{NIR}}}-\pi)\sin(\pi t/(4T_{\textrm{\textrm{NIR}}}))^{2}$
with $T_{\textrm{\textrm{NIR}}}=110.32$ a.u. ($=2.6685$fs). This
corresponds to a wavelength of $800$nm, and a total duration of $10.67$fs.
The peak amplitude is $A_{\textrm{\textrm{NIR}}}=0.6627$ (corresponding
to an intensity of $5\times10^{13}\textrm{W/cm}^{2}$). The XUV pulses
have the form: $u_{\textrm{XUV}}=A_{\textrm{XUV}}\sin(2\pi t/T_{\textrm{XUV}})\exp(-(t-5T_{\textrm{XUV}}/4)^{2}/(2\sigma^{2}))$
with $T_{\textrm{XUV}}=2.07$a.u. which corresponds to a wavelength
of $15$nm. We take $\sigma=6.207$a.u. ($150$ attoseconds) and peak
amplitude $A_{\textrm{XUV}}=0.08$a.u. The control strategy is to
generate a sequential double ionization with the two successive XUV
pulses at the peaks of the NIR pulse.

For this simulation the PvB calculation is based on an underlying
grid with a range $x\in[-400;400]$ with $N=4000$ points in each
dimension. On such a grid, the dimension of the unreduced Hilbert
space is $16\times10^{6}$ while the maximum dimension of the reduced
Hilbert space used during the dynamics is $28207$. This translates
to a reduction by a factor of $400$ in the size of the Hilbert space and hence five orders of magnitude in the number of elements
in the Hamiltonian, as compared to the size of the unreduced Hamiltonian.

\begin{figure}
\noindent \begin{centering}
\includegraphics[scale=0.5]{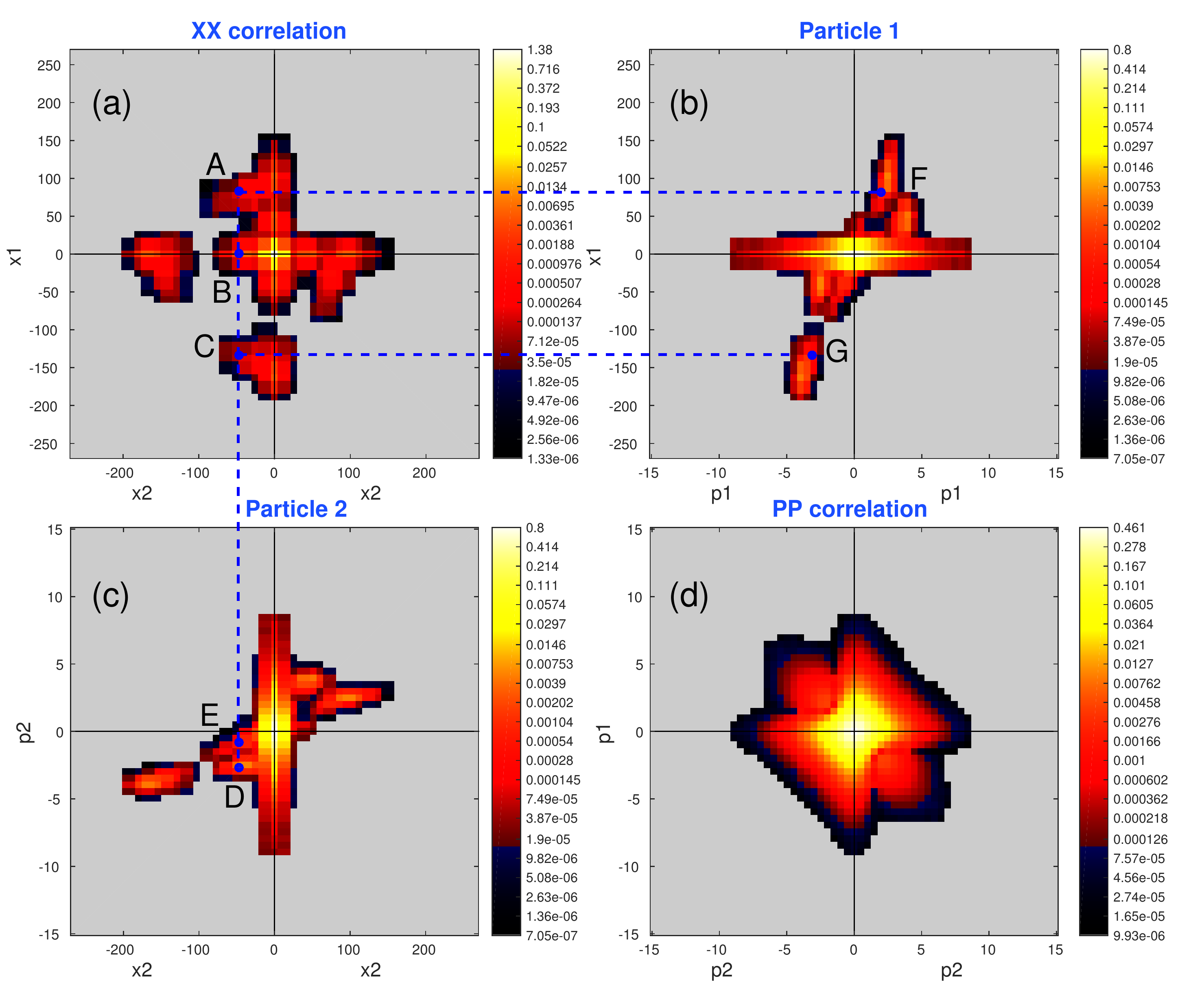}
\par\end{centering}

\noindent \centering{}\protect\caption{\label{fig:He1D_phasespace_275}Phase space state of the Helium two electron wavefunction after the second XUV pulse at $t=275$a.u (third red circle in fig. \ref{fig:H1D_TDSE_progression}). Interpretation of simulation results is that events unfold as follows: A single ionization is generated by the first XUV pulse, and is denoted by (F) and (G) in the Particle 1 panel, top-right. Then two  double ionization event are triggered by both the second XUV pulse and the on-going NIR pulse, (A) and (C) in the XX correlation panel, top-left.  The contributions of the two pulses are distinguishable in the single-particle projection (bottom left panel), with the second XUV pulse generating the (D) ionization, and the NIR generating the (E) ionization. The second XUV pulse also triggers a second single ionization event, denoted by (B). For an in depth analysis, see  \cite{PvB-MCTDH-comparison-paper}.
}
\end{figure}

For the results of the simulation, see fig. \ref{fig:H1D_TDSE_progression}
and \ref{fig:He1D_phasespace_275}. The two-particle phase space is
four dimensional, and therefore we display 2D projections. In fig. \ref{fig:H1D_TDSE_progression} we show the evolution of the
two-electron system as a function of time at three times: just
prior to the second XUV pulse, just after the second XUV pulse and
later when the wavepackets have had sufficient time to drift away
from the core.

In figure \ref{fig:He1D_phasespace_275} we look in depth at the state
of the system corresponding to the bottom row of figure \ref{fig:H1D_TDSE_progression}.
The 4D phase space is projected into $4$ planes: $x_{1},x_{2}$ and
$p_{1},p_{2}$ as well as the single particle projections $x_{1},p_{1}$
and $x_{2},p_{2}$. The wavefunction exhibits complex contributions,
which may be associated with specific NIR and XUV pulses. One
may observe interesting phenomena such as both sequential and concerted double
ionization. For an in depth discussion of these results, refer to
 \cite{PvB-MCTDH-comparison-paper}.


\section{\label{sec:Conclusions-and-outlook}Conclusions and outlook}

We presented the Periodic von Neumann method with Biorthogonal exchange (PvB) for quantum
mechanical simulations, based on the von Neumann lattice of phase space Gaussians.
The use of periodic boundary conditions allows the method to converge with Fourier accuracy,
while the exchange of the role of the basis and its biorthogonal functions allows the elimination of basis functions
that are not in the occupied regions of phase space.
This provides a sparse representation, leading to a significant savings in memory
and CPU resources.

The PvB approach utilizes a phase space-local non-orthogonal basis,
which is related to an underlying Fourier grid via a similarity transformation.
By carefully defining the projection operator for non-orthogonal bases,
we were able to project out all the unoccupied areas of phase space,
leaving a far smaller subspace to contend with, while maintaining
a very straightforward control of accuracy.

Appropriate algorithms for solving the TISE and TDSE were developed.
In both cases one must contend with the a priori ignorance of the
exact phase space area needed to represent the final state. For finding
eigenmodes, we iteratively grow the area from "seeds" at the local minima of the potential,
expanding and pruning the area as needed. For dynamics,
we make use of the smooth transition of the state through phase space to
keep track of the changes in the occupied phase space as a function of time.

The method was demonstrated
on the textbook example of the double-well potential, as well as the
more demanding simulation of the double-ionization of helium in one-dimension.

While PvB is now a mature method, with mathematical rigor and
fully fleshed-out algorithms, significant further development is possible.
Additional approximations may be introduced, beyond projecting out
areas of phase space where the wavefunction amplitude is below the
specified threshold. Specifically, one may further reduce the representation
by decomposing multi-dimensional objects into a sum-of-products series,
and truncating the series when the correlation is sufficiently low,
\'{a} la POTFIT (see Section \ref{sub:Decomposition-POTFIT}). While
this decomposition is currently utilized in the representation of the unreduced
Hamiltonian, much will be gained if this approach can be extended
to the reduced state and the reduced Hamiltonian. This is a challenging
problem, however, as the dynamics continuously modify the reduced
basis, which is generally not easily decomposable.

Further areas of research include the correspondence between PvB and
other phase space representations, such as the Wigner representation.
We anticipate that PvB will be significantly more efficient because of its intrinsically discrete representation and its excellent convergence properties. We also plan to explore the explicit incorporation of particle
symmetries in multi-particle systems (bosonic, fermionic) into the
PvB framework.

Beyond method development, a wide range of problems in atomic and molecular physics
can be addressed using the PvB methodology, from electron dynamics in multi-electron systems to nuclear dynamics in polyatomic molecules. We are currently exploring these applications.

To conclude, PvB is an accurate, well-controlled, scalable and efficient
method for quantum dynamics simulations. It is our hope it will
find its place as part of the standard quantum numerics toolbox.

\section*{Acknowledgement}

We wish to thank Henrik Larsson and Tucker Carrington for enlightening conversations. This work was supported
by the Israel Science Foundation (533/12), the Minerva Foundation with funding from the Federal German Ministry for Education and Research and the Koshland Center for Basic Research.

\vspace{15mm}
\section*{Bibliography}


\end{document}